\documentclass[prl,showpacs,aps,twocolumn,preprintnumbers]{revtex4-2}% Physical Review Letters

\usepackage{amssymb}
\usepackage{amsmath}
\usepackage{graphicx}% Include figure files
\usepackage{float}% Package added by Dan 06/06/2020 - For image floats, which are used for positioning
\usepackage{subcaption}% Package added by Dan 06/06/2020 - For subfigures
\usepackage{bm}% bold math

\usepackage[letterpaper, margin=0.75in]{geometry}

\usepackage{caption}
\begin{document}

\title{Two-Orders-of-Magnitude Improvement in the Total Spin Angular Momentum of $^{131}$\/Xe Nuclei Using Spin Exchange Optical Pumping}

\author{Michael J. Molway}
\affiliation{%
Department of Chemistry and Biochemistry, Southern Illinois University, Carbondale, IL 62901, USA
}%

\author{Liana Bales-Shaffer}
\affiliation{%
Department of Chemistry and Biochemistry, Southern Illinois University, Carbondale, IL 62901, USA
}%

\author{Kaili Ranta}
\altaffiliation[Current address: ]{School of Medicine, The Ohio State University, Columbus, OH, USA}
\affiliation{%
Department of Chemistry and Biochemistry, Southern Illinois University, Carbondale, IL 62901, USA
}%

\author{Dustin Basler}
\affiliation{%
Department of Chemistry and Biochemistry, Southern Illinois University, Carbondale, IL 62901, USA
}%

\author{Megan Murphy}
\affiliation{%
Department of Chemistry and Biochemistry, Southern Illinois University, Carbondale, IL 62901, USA
}%

\author{Bryce E. Kidd}
\affiliation{%
Department of Chemistry and Biochemistry, Southern Illinois University, Carbondale, IL 62901, USA
}%

\author{Abdulbasit Tobi Gafar}
\affiliation{%
Department of Chemistry and Biochemistry, Southern Illinois University, Carbondale, IL 62901, USA
}%

\author{Justin Porter}
\affiliation{%
Department of Chemistry and Biochemistry, Southern Illinois University, Carbondale, IL 62901, USA
}%

\author{Kierstyn Albin}
\affiliation{%
Department of Chemistry and Biochemistry, Southern Illinois University, Carbondale, IL 62901, USA
}%

\author{Boyd M. Goodson}
 \email{bgoodson@chem.siu.edu}
\affiliation{%
Department of Chemistry and Biochemistry, Southern Illinois University, Carbondale, IL 62901, USA
}%

\author{Eduard Y. Chekmenev}
\affiliation{
Department of Chemistry, Integrative Biosciences (Ibio), Karmanos Cancer Institute (KCI), Wayne State University, Detroit, Michigan 48202, USA, 
Russian Academy of Sciences, 
Leninskiy Prospekt 14, Moscow, 119991, Russia
}%

\author{Matthew S. Rosen}
\affiliation{
A.A. Martinos Center for Biomedical Imaging, Massachusetts General Hospital and Harvard Medical School, Boston, MA 02129; Department of Physics, Harvard University, Cambridge, MA 02138, USA
}%

\author{W. Michael Snow}
\affiliation{
Department of Physics, Indiana University, Bloomington, IN, USA
}%

\author{James Ball}
\affiliation{
School of Medicine, University of Nottingham, Queens Medical Centre, Nottingham, UK
}%

\author{Eleanor Sparling}
\affiliation{
School of Medicine, University of Nottingham, Queens Medical Centre, Nottingham, UK
}%

\author{Mia Prince}
\affiliation{
School of Medicine, University of Nottingham, Queens Medical Centre, Nottingham, UK
}%

\author{Daniel Cocking}
\affiliation{
School of Medicine, University of Nottingham, Queens Medical Centre, Nottingham, UK
}%

\author{Michael J. Barlow}
\affiliation{
School of Medicine, University of Nottingham, Queens Medical Centre, Nottingham, UK
}%

\date{\today}

\begin{abstract}
 We report on hyperpolarization of quadrupolar ($I=3/2$) $^{131}$\/Xe via spin-exchange optical pumping. Observations of the $^{131}$\/Xe polarization dynamics show that the effective alkali-metal/$^{131}$\/Xe spin-exchange cross-sections are large enough to compete with $^{131}$\/Xe spin relaxation.  $^{131}$\/Xe polarization up to 7.6\/$\pm$\/1.5$\%$ was achieved in $\sim$\/8.5$\times10^{20}$ spins---a $\sim$\/100-fold improvement in the total spin angular momentum---enabling applications including measurement of spin-dependent neutron-$^{131}$\/Xe s-wave scattering and sensitive searches for time-reversal violation in neutron-$^{131}$\/Xe interactions beyond the Standard Model.

%Valid PACS numbers may be entered using the \verb+\pacs{#1}+ command.
\end{abstract}

%82.56.-b  NMR in chemical physics
%32.80.Xx  optical pumping of atoms
%87.61.-c  magnetic resonance imaging in medical physics
%42.55.Px  diode lasers 

% may need to add nuclear physics PACS codes?
%\pacs{32.80.Xx, 82.56.-b, 87.61.-c, 42.55.Px}% PACS, the Physics and Astronomy
                             % Classification Scheme.

\maketitle
% Had to add in the usepackage{graphicx} for this to work.

Spin exchange optical pumping (SEOP)~\cite{WalkerHapper} can polarize macroscopic amounts of $I$\/=\/$1/2$ nuclei like $^{3}$He and $^{129}$\/Xe~\cite{bascontflow,rosen1999,zook,meersmann129,hersmanPRL,whitingJMR,sheffield}  to near-unity values~\cite{NikolaouhighP2,NikolaouhighP} for applications in biomedical imaging, NMR spectroscopy, and fluid dynamics \cite{boydrev2002,lungrev,danillarev}, as well as nuclear/particle physics efforts to constrain CPT/Lorentz violations~\cite{bear2000,cane2004,gemmel2010,Allmendinger2014,Stadnik2015,Kostelecky2018} and electric dipole moment searches~\cite{Sachdeva2019}. However, the strong quadrupole moment of higher-spin nuclei like $I$\/=\/$3/2$\/ $^{131}$\/Xe gives rise to much faster spin-lattice relaxation ($T_1$\/) losses with increasing Xe density and surface interactions \cite{Brinkmann1962,Butscher1994,Stupic2011}. This physics has severely constrained the utility of $^{131}$\/Xe for many applications, even when employing small amounts ($\lesssim$\/10$^{16}$ spins) of polarized ground state $^{131}$\/Xe nuclei~\cite{Volk1980,Kwon1981,Wu1987,Wu1988,Butscher1994,mehringgyro,Raftery1994,luo1999,walkerNsnow,Vershovskii131,petrov131,Walker1312019} to understand the atomic physics of the SEOP process~\cite{Volk1980,Kwon1981}, probe gas/surface interactions~\cite{Wu1987,Wu1988,Butscher1994}, study Berry geometric phases~\cite{mehringgyro}, or perform $^{129}$\/Xe/$^{131}$\/Xe comagnetometry~\cite{walkerNsnow,Vershovskii131,petrov131,Walker1312019}---including searches for axion-like particles~\cite{walkerNsnow}; minute amounts of polarized $^{131m}$\/Xe have also been used to demonstrate a new MRI/gamma-ray imaging modality~\cite{meta131nature}. 
 While larger amounts of hyperpolarized (HP) $^{131}$\/Xe may be produced transiently via cross-polarization with HP $^{129}$\/Xe in cryogenic lattices~\cite{gatzke1993}, arguably the greatest success with bulk production of HP $^{131}$\/Xe to date~\cite{Stupic2011} was part of pioneering work by Meersmann and co-workers to hyperpolarize quadrupolar noble gas nuclei in the gas-phase via SEOP \cite{Pavlovskaya2005,Hughes2013,lilburn2013,six2014}. They reported $^{131}$\/Xe polarizations of 2.2\%, 0.44\%, and 0.026\% at 9.4 T for 0.075, 0.3, and 1.4 bar Xe, respectively, following gas separation from a $\sim$\/72 mL SEOP cell~\cite{Stupic2011}. 
The demonstration of a significant improvement in the product $P_{Xe}N$, where $P_{Xe}$ is the nuclear polarization and $N$ is the number of polarized $^{131}$\/Xe nuclei, would be of scientific interest to a variety of areas in science and technology. 

Here we investigate the preparation of hyperpolarized $^{131}$\/Xe using isotopic enrichment, next-generation spectrally-narrowed laser diode arrays, and {\it in situ} low-field NMR to permit real-time observation of polarization dynamics and SEOP optimization. Our extraction of effective alkali-metal / $^{131}$\/Xe spin-exchange cross-sections indicates that the resulting spin-exchange rates can be large enough to compete with rapid $^{131}$\/Xe spin relaxation in glass cells. $^{131}$\/Xe polarization values as high as 7.6$\%$\/$\pm$\/1.5$\%$ were achieved at 0.37 amagat \cite{amagatdef} in a ~0.1 L cell ($N=8.5\times10^{20}$ $^{131}$\/Xe spins).  This $\sim$\/100-fold improvement in $P_{Xe}N$ for $^{131}$Xe nuclei has already led to a new scientific application: the first measurement of the spin-dependent scattering length of polarized neutrons from polarized $^{131}$Xe nuclei~\cite{Lu}. We conclude this paper with a brief discussion of future scientific applications, including searches for time-reversal invariance violation (TRIV) in polarized neutron transmission at the 3.2 eV p-wave neutron-nucleus resonance in polarized $^{131}$\/Xe. The sensitivity of such a search can discover new TRIV sources beyond the Standard Model.   

Our SEOP apparatus has been described elsewhere~\cite{whitingJMR,saha}. Briefly, a circularly-polarized, spectrally-narrowed laser tuned to the Rb or Cs D1 line (794.8 nm or 894.3 nm) enters a cylindrical Pyrex cell ($V$\/$\approx$\/0.1 L) loaded with Rb or Cs, 300 torr enriched (84\%) $^{131}$\/Xe, and 300 torr N$_2$. The cell resides in a forced-air oven~\cite{nikolaou_JACS_2014} within a Helmholtz coil (HC) pair electromagnet. Additional details are in the Supplemental Information (SI).

\begin{figure}[h]
    \centering
    \includegraphics[width=0.5\textwidth]{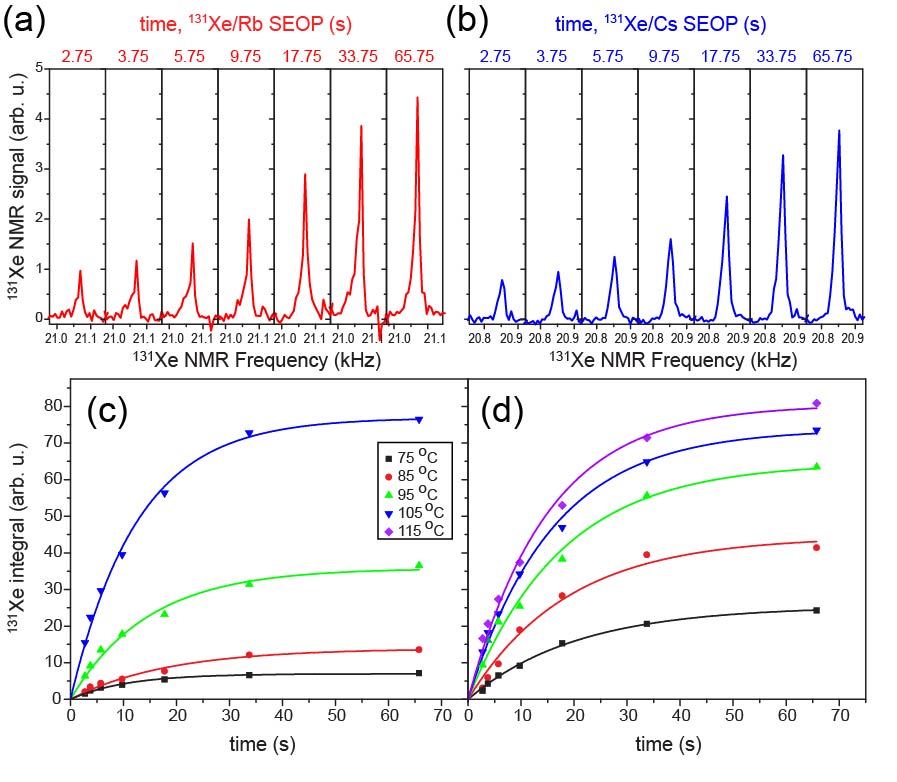}
   \caption{(a,b) Single-scan {\it in situ} low-field $^{131}$\/Xe NMR spectra acquired at 20.9 kHz, 105 $^{\circ}$\/C at each delay time ($\tau$\/) for Rb (a) and Cs (b). (c,d) $^{131}$\/Xe polarization dynamics curves from SEOP with Rb (c) and Cs (d) generated by fitting the integrated $^{131}$\/Xe NMR signal (average of 24 scans, see SI) at each $\tau$\/ to the exponential function: $S(t)=S_0 [1-\mathrm{exp}(-t\Gamma_{SEOP})]$.}
    \label{fig:pol_dynamics}
\end{figure}

Figures~\ref{fig:pol_dynamics}(a,b) show time-dependent {\it in situ} $^{131}$\/Xe NMR spectra during SEOP with Rb (a) or Cs (b). Use of isotopically enriched $^{131}$\/Xe and high resonant photon fluxes enabled single-shot $^{131}$\/Xe detection, improving on previous work~\cite{kailidiss} and permitting optimization of $^{131}$\/Xe SEOP and real-time monitoring of $^{131}$\/Xe polarization dynamics as a function of temperature for each metal (Figs. \ref{fig:pol_dynamics}(c,d)). The signals were fit to $S(t)=S_0 [1-\mathrm{exp}(-t\Gamma_{SEOP})]$ to extract the polarization rate constant for each curve, $\Gamma_{SEOP}$\/.

\begin{figure}[h]
    \centering
    \includegraphics[width=0.45\textwidth]{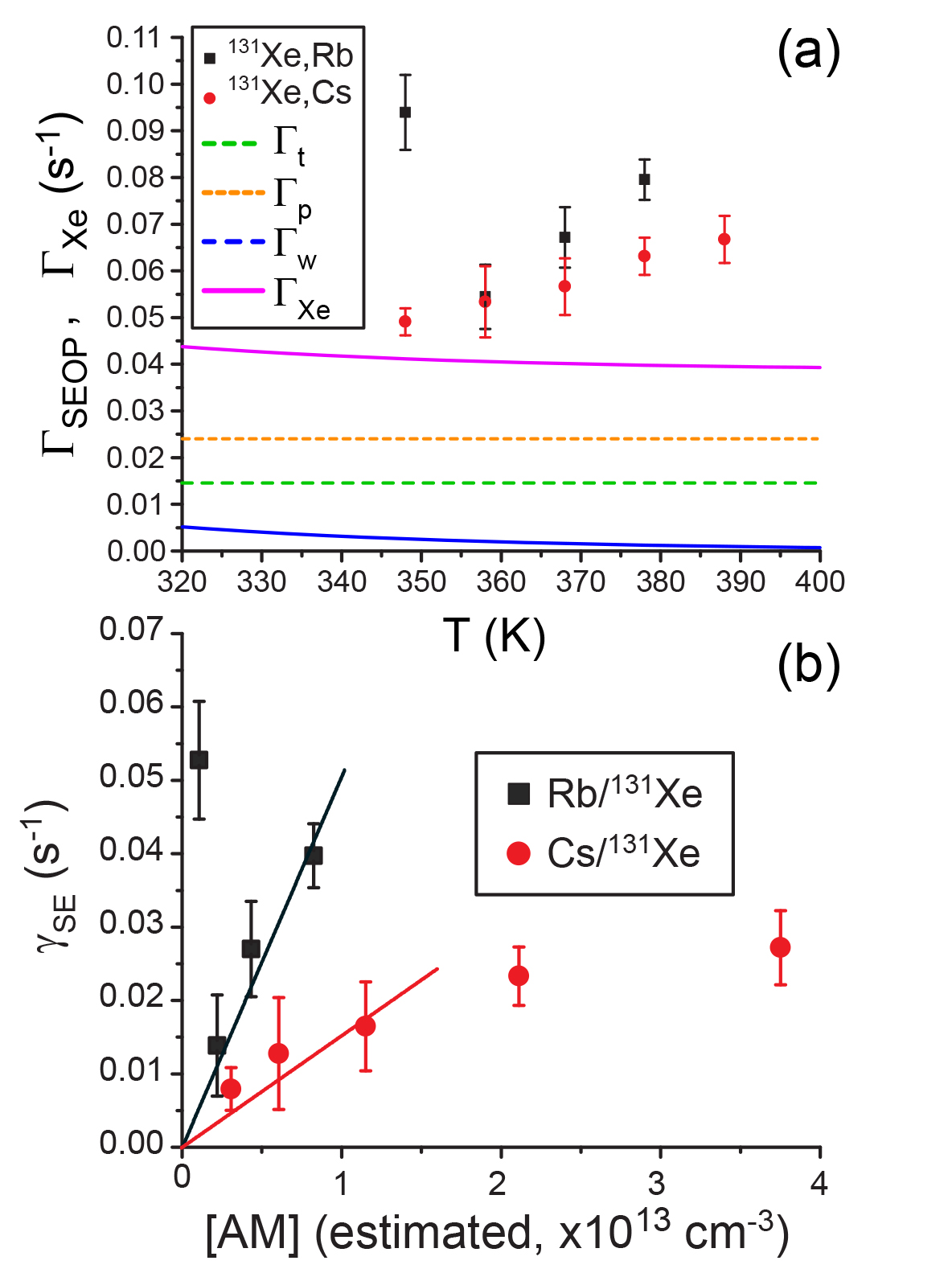}
   \caption{(a) $^{131}$Xe SEOP rate constants, $\Gamma_{SEOP}$, versus temperature for both Rb/Xe and Cs/Xe; estimated $^{131}$Xe spin relaxation contributions are plotted for comparison (see SI). Error bars are fit uncertainties. (b) $^{131}$Xe/Rb and $^{131}$Xe/Cs spin-exchange rates ($\gamma_{SE}$\/), after subtraction of the estimated contributions from $\Gamma_{Xe}(131)$\/ versus estimated alkali metal vapor densities~\cite{SteckCsRb}. Lines are fits (using three points each, selected over similar ranges of estimated [AM] values) forced through zero (i.e. extrapolating to [AM]=0) to yield estimated $\gamma^{\prime}$\/ values. For Cs: 1.5($\pm$0.2)$\times10^{-15}$\/ cm$^{3}\cdot$\/s$^{-1}$, $R^2$\/=0.967; for Rb: 5.1($\pm$0.4)$\times10^{-15}$\/ cm$^{3}\cdot$\/s$^{-1}$, $R^2$\/=0.988.}
    \label{fig:polrelaxation}
\end{figure}

Resulting $\Gamma_{SEOP}$\/ values (Fig.~\ref{fig:polrelaxation}(a)) have contributions from alkali-metal / $^{131}$\/Xe spin-exchange and $^{131}$\/Xe nuclear spin relaxation: $\Gamma_{SEOP}$\/=$\gamma_{SE}$\/+$\Gamma_{Xe}$\/(131), where $\gamma_{SE}$ is the spin-exchange rate and $\Gamma_{Xe}$\/(131)=$1/T_1$\/($^{131}$Xe). As with $^{129}$\/Xe, higher alkali metal densities (i.e. higher cell temperatures) increase $^{131}$\/Xe polarization rates according to the relation $\gamma_{SE}$\/=$\gamma^{\prime}$\/[Rb], where $\gamma^{\prime}$ is the per-atom spin-exchange rate (or effective spin-exchange cross section). As pointed out by Volk and co-workers~\cite{Volk1980,Kwon1981}, one can test the theory of Herman~\cite{Herman1965}---developed to calculate contributions to Rb/gas spin exchange from scalar versus classical dipolar hyperfine interactions---by using it to predict the ratio of spin-exchange cross sections for isotopes of the same element with atoms of a given metal vapor by measuring the relative sizes of $\gamma^{\prime}$ for $^{131}$\/Xe and $^{129}$\/Xe:

\begin{equation}
\frac{\gamma^{\prime}_{I}}{\gamma^{\prime}_{K}} = \frac{I(I+1)\gamma_I^2}{K(K+1)\gamma_K^2},
    \label{eq:ratioofcross}
\end{equation}

\noindent where $I$, $K$ are the nuclear spins and $\gamma_{I(K)}$ are the corresponding gyromagnetic ratios. For $^{129}$\/Xe and $^{131}$\/Xe this equation predicts a ratio of $\approx$\/2.28.  Volk and co-workers reported a $\sim$\/16-fold smaller $\gamma^{\prime}$\/ value for Rb/$^{131}$\/Xe than that measured for Rb/$^{129}$\/Xe \cite{Volk1980}.  However, later experiments suggested somewhat closer agreement with the theoretical prediction (e.g., corresponding to a  $\sim$\/4-8-fold ratio \cite{Kwon1981, Butscher1994}). The $^{131}$\/Xe $\gamma_{SE}$ is practically important for the envisioned TRIV experiments:
Given the much greater relaxation rates for $^{131}$\/Xe versus $^{129}$\/Xe, a large concomitant reduction in $\gamma_{SE}$ could impact the feasibility of bulk hyperpolarized $^{131}$\/Xe production by SEOP.

Extracting $\gamma_{SE}$ values from $^{131}$\/Xe polarization dynamics is confounded by the inability to independently measure $\Gamma_{Xe}$\/(131) and its large size relative to $\gamma_{SE}$ (compared to the case of $^{129}$\/Xe SEOP). Guided by previous efforts \cite{Brinkmann1962,Stupic2011,Butscher1994,Volk1980,Kwon1981,Wu1987,Wu1988,Raftery1994,adrian1965,Chann2002,Anger2008}), we constructed a model of gas-phase $^{131}$\/Xe relaxation (see SI).  Briefly,  $^{131}$\/Xe relaxation may be broken down as: $\Gamma_{Xe}(131)$\/=$\Gamma_t$\/+$\Gamma_p$\/+$\Gamma_w$, where
$\Gamma_t$\/=$\alpha \mathrm{[Xe]}$ arises from transient (binary) Xe-Xe collisions (where  $\alpha$\/=3.96$\times10^{-2}$\/ amg$^{-1}\cdot$\/s$^{-1}$ \cite{Brinkmann1962,amagatdef}), $\Gamma_p$\/=$\beta \chi_{\mathrm{Xe}}'$ is from persistent Xe-Xe dimers (where $\beta$\/=0.036 s$^{-1}$\/ and $\chi_{\mathrm{Xe}}'$ = [Xe]$\cdot$\/([Xe]+0.51[N$_2$])$^{-1}$), and $\Gamma_w$ arises from wall collisions. Extrapolation of previous measurements \cite{Brinkmann1962,Stupic2011,Butscher1994} to our conditions allowed the contributions to $\Gamma_{Xe}(131)$\/ to be estimated (Fig. \ref{fig:polrelaxation}(a)); subtraction of $\Gamma_{Xe}(131)$\/ from the $\Gamma_{SEOP}$\/ values in Fig. \ref{fig:polrelaxation}(a) permitted $\gamma_{SE}$ estimates to be plotted against [AM] values predicted from vapor-pressure curves \cite{SteckCsRb} (Fig. \ref{fig:polrelaxation}(b)). 

Although alkali metal densities can differ from vapor pressure curve predictions~\cite{jau3papers,hughes2005,chann2002PRA}, a linear fit of the $\gamma_{SE}$ data can provide an estimate of $\gamma^{\prime}$\/ for each alkali metal / $^{131}$\/Xe combination (see SI). Fitting the Rb data while extrapolating [Rb] to zero (and excluding the anomalous first point) gives $\gamma^{\prime}$\/(Rb/$^{131}$\/Xe)=5.1($\pm$0.4)$\times10^{-15}$\/ cm$^{3}\cdot$\/s$^{-1}$\/. To fit Cs over a similar range of [AM] and $\Delta T$\/, the first three points were fit (extrapolating [Cs] to 0) to give $\gamma^{\prime}$\/(Cs/$^{131}$\/Xe)=1.5($\pm$0.2)$\times10^{-15}$\/ cm$^{3}\cdot$\/s$^{-1}$. Our previous measurements of Rb/$^{129}$\/Xe and Cs/$^{129}$\/Xe spin exchange with roughly similar conditions obtained  $\gamma^{\prime}$\/(Rb/$^{129}$\/Xe)=1.67($\pm$0.06)$\times10^{-15}$\/ cm$^{3}\cdot$\/s$^{-1}$ and $\gamma^{\prime}$\/(Cs/$^{129}$\/Xe)=2.6($\pm$0.1)$\times10^{-15}$\/ cm$^{3}\cdot$\/s$^{-1}$, respectively \cite{nick2011}. 
Considering Cs/Xe first, these results would yield a $\gamma^{\prime}$\/(129):$\gamma^{\prime}$\/(131) ratio of 1.7$\pm$\/0.3---in reasonable agreement with the theory of Herman~\cite{Herman1965}. Moreover, albeit under different conditions, comparisons of spin-exchange values of Hughes and coworkers~\cite{hughes2005} with the above $\gamma^{\prime}$\/(Cs/$^{131}$\/Xe) value would correspond to ratios of $\sim$1.3-3.6---again in rough agreement with  Herman. However, the value for $\gamma^{\prime}$\/(Rb/$^{131}$\/Xe) is significantly {\it larger} than the corresponding $\gamma^{\prime}$\/(Rb/$^{129}$\/Xe) values reported in Refs.~\cite{nick2011,hughes2005}; moreover, in those works, a $\sim$\/1.5-2-fold {\it smaller} $\gamma^{\prime}$ value was observed for Rb/$^{129}$\/Xe compared to Cs/$^{129}$\/Xe, and it is unclear why the trend would be reversed for $^{131}$\/Xe. One explanation may be that differences in laser excitation (and corresponding AM heating, see SI) used here for Rb versus Cs gave rise to higher-than-expected [Rb] values, causing  $\gamma^{\prime}$\/(Rb/$^{131}$\/Xe) to be overestimated. Regardless, the $^{131}$\/Xe SE cross-sections for both Cs and Rb appear to be high enough to compete with $^{131}$\/Xe auto-relaxation and allow significant $^{131}$\/Xe $P_{Xe}$\/ values---even at relatively high Xe densities.  

Determining the absolute $^{131}$\/Xe polarization $P_{Xe}$ from this first dataset was confounded by the large difference in gyromagnetic ratios (i.e. frequencies) for $^{131}$\/Xe versus the thermally polarized reference ($^1$\/H in water). Indirect comparisons through a third nucleus (HP $^{129}$\/Xe) allowed $P_{Xe}$ to be initially estimated at $\sim$\/1-10\%.    

\begin{figure}[h]
    \centering
    \includegraphics[width=0.5\textwidth]{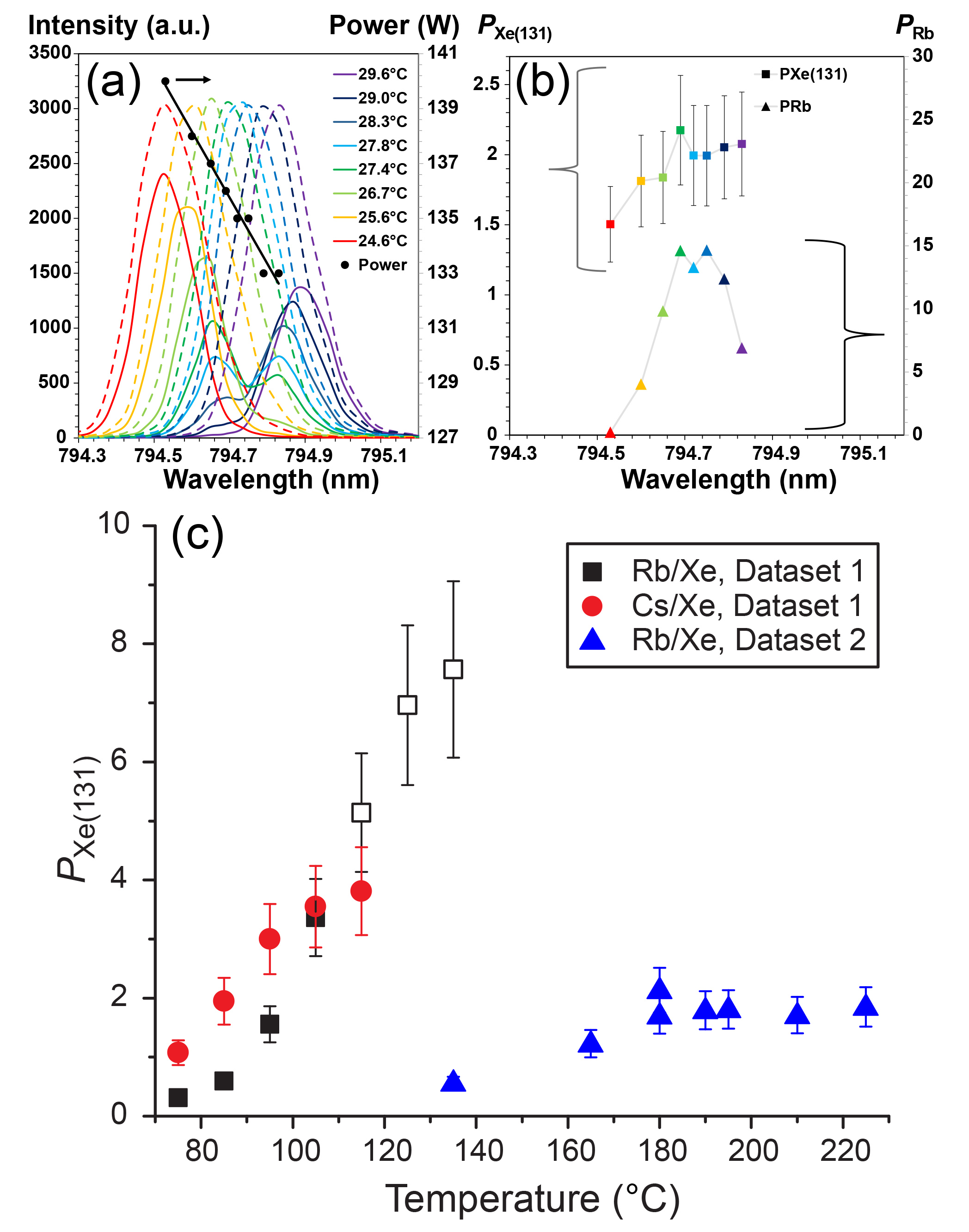}
   \caption{(a) Near-IR spectra for the SEOP laser transmitted through the Rb cell when at room temperature (dotted) and at 180 $^{\circ}$C (with magnet on, solid lines) as the laser is tuned by varying the LDA temperature, as well as corresponding power measurements (black circles / line). (b) $P_{Xe}$ (squares) and Rb polarization values (triangles, estimated from laser transmission variations with field cycling \cite{whitingJMR}) for each laser centroid wavelength. (c) Absolute $P_{Xe}$ values as a function of cell temperature for both datasets; open icons are single-point acquisitions taken under conditions of ``Rb runaway."}
    \label{131pol_v_wave_n_T}
\end{figure}

To measure $P_{Xe}$, a second set of experiments was performed using the Rb cell. First, measurements of the $^{131}$\/Xe NMR signal as a function of laser wavelength found that the polarization was generally higher when the laser centroid was red-shifted from the Rb D1 absorption maximum (Figs.~\ref{131pol_v_wave_n_T}(a,b))---consistent with previous $^{129}$\/Xe results when using a spectrally narrowed laser in the power-limited regime \cite{VHGonchip}.  Lower-than-expected $^{131}$\/Xe signals---despite the higher RF frequency (46 kHz) and elevated cell temperatures compared to the first dataset---were found to result from Rb loss from the cell's optical region, giving greatly reduced spin-exchange rates (see SI).  The cell was then washed and used as a secondary thermally polarized water reference, allowing its matched geometry to reduce systematic errors in $P_{Xe}$\/. The ability to detect NMR signals from HP $^{131}$\/Xe and $^1$\/H from thermally polarized water at the same frequency (at $\sim$\/13 and $\sim$\/1.1 mT, respectively) allowed $P_{Xe}$ to be meaasured as a function of offset wavelength (Fig. \ref{131pol_v_wave_n_T}(b)) and cell temperature (Fig. \ref{131pol_v_wave_n_T}(c)), with $P_{Xe}$\/ peaking at 2.17$\%$\/$\pm$\/0.39$\%$ for this second dataset. Measurements of other relevant systematics (see SI) determined $P_{Xe}$ values from the first dataset (Fig.~\ref{131pol_v_wave_n_T}(c)).  Under conditions for the dynamics experiments in Fig.~\ref{fig:pol_dynamics}, $P_{Xe}$ values of up to 3.4$\pm$\/0.7\% and 3.8$\pm$\/0.7\% were measured for Rb/Xe and Cs/Xe SEOP, respectively. Even greater values were measured under higher-temperature conditions of ``Rb runaway" (\cite{zook,witte,nikolaou_maps_2014} wherein laser heating causes positive feedback with Rb density), reaching values as high as $P_{Xe}$\/=\/7.6$\%$\/$\pm$\/1.5$\%$. Employing a metal heating jacket should enable more stable SEOP control at high cell temperatures~\cite{jacket}.

The large improvement in the $^{131}$Xe polarization-density product achieved in this work has already enabled the first measurement of the spin-dependent neutron-$^{131}$\/Xe s-wave scattering amplitude~\cite{Lu} using neutron pseudomagnetic precession~\cite{Baryshevsky1965, Forte1973,  Abragam1973, Abragam1982, Pokazanev1979,  Zimmer2002, Tsulaia2014}. The same techniques can be applied to measure neutron pseudomagnetic precession in resonance scattering to compare with a recently-developed theory~\cite{Gudkov2017} using an instrument at the LANSCE neutron source at Los Alamos~\cite{Schaper2020}. Measurements in progress similar to those described in~\cite{Okudaira2018} will determine the neutron resonance mixing spectroscopic parameter $\kappa$, which relates the parity-odd, time-odd  (P-odd, T-odd) cross section difference $\Delta \sigma_{PT}$ and the P-odd cross section difference $\Delta \sigma_{P}$~\cite{Gudkov2018}. A neutron-$^{131}$\/Xe time reversal violation search is also sensitive to axion-like particle (ALP) exchange~\cite{Fadeev2019}, which is poorly constrained by electric dipole moment (EDM) searches for ALP masses above 10 meV~\cite{Dzuba2018} as in this case the Standard Model axion relation between axion mass and coupling constant does not apply~\cite{Mantry2014}. This work enables a direct ALP search with sensitivity surpassing indirect EDM constraints for ALP masses in the eV-MeV region using SEOP-polarized $^{131}$Xe. 

Sensitive searches for new sources of TRIV in strongly interacting systems can address fundamental questions in cosmology and particle physics. If Sakharov's arguments~\cite{Sak67} for the origin of the matter-antimatter asymmetry of the universe in Big Bang cosmology are correct, then new sources of time-reversal violation await discovery. A discovery of TRIV in nucleon-nucleus interactions 
would also indicate new physics beyond the Standard Model of particles and interactions. Searches for EDMs of elementary systems are a sensitive method for TRIV investigation~\cite{Chupp2019}. Another sensitive method, which has been proposed but not yet realized experimentally, would employ transmission of polarized neutrons passing through polarized nuclei with p-wave resonances that are known to greatly amplify P-odd asymmetries by $\sim$\/5‐6 orders of magnitude~\cite{Mitchell1999, Mitchell2001}. Since the amplification mechanism in these neutron resonances works for any P-odd interaction, any TRIV effects that violate P will also be amplified by a similar amount. A P-odd, T-odd search in polarized neutron transmission would search for a term in the forward scattering amplitude of the form $\vec{s} \cdot [\vec{k} \times \vec{I}]$ where $\vec{k}$ is the neutron momentum, $\vec{s}$ is the neutron spin, and $\vec{I}$ is the nuclear polarization. This experiment would be realized as a null-test for TRIV~\cite{Bowman2014} similar to EDM searches. Since these two methods possess different sensitivities to the many different possibilities for P and T violation in nucleons, any qualitative advance that enables a sensitive TRIV search in polarized neutron transmission is of general interest to the scientific community.

$^{131}$\/Xe is a promising nucleus for a polarized neutron optics TRIV search. It possess one of the largest P-odd analyzing powers in neutron-nucleus resonance reactions, 4.3$\%$\/ on the 3.2 eV p-wave resonance~\cite{Szymanski1996,skoy1996}. The spatially uniform density and polarization of nuclei polarized with spin-exchange optical pumping (SEOP) helps suppress systematic errors in polarized neutron transmission. Moreover, $^{129}$\/Xe can be polarized in the same cell along with $^{131}$\/Xe to help suppress other potential systematic errors associated with the spin dependence of the neutron-nucleus strong interaction. These extra handles on systematic errors may be crucial for an experiment of this type, which has never been conducted.  However, until now the figure of merit governing the sensitivity of a neutron optics TRIV experiment, the polarization-number product $P_{Xe}N$, has been too small to mount an experiment with a scientifically-interesting sensitivity. 

In conclusion, we have optimized SEOP to polarize $8.5\times10^{20}$ $^{131}$\/Xe nuclei to 7.6$\%$\/$\pm$\/1.5$\%$, a $\gtrsim$\/$10^{2}$\/-fold improvement in $P_{Xe}N$ over previous efforts. $P_{Xe}N$ is large enough to conduct a sensitive search for TRIV in neutron-$^{131}$\/Xe scattering and has already enabled the first measurement of neutron-$^{131}$\/Xe pseudomagnetic precession. These results may inspire further improvements in SEOP polarization of other quadrupolar noble gas isotopes (e.g. $^{83}$\/Kr \cite{six2014} and $^{21}$\/Ne \cite{romalis21ne}), with applications ranging from medical imaging to searches for CPT/Lorentz violation. Higher $P_{Xe}N$ using more laser power delivered over larger volumes will benefit neutron TRIV searches and may enable pulmonary imaging with HP $^{131}$\/Xe, with its surface-selective relaxation potentially providing orthogonal contrast to that of HP $^{129}$\/Xe as shown for HP $^{83}$\/Kr \cite{six2014}.

We acknowledge funding support from NSF (CHE-1905341, CHE-1904780, and REU funding from the NSF/DoD ASSURE Program, DMR-1757954), DoD CDMRP (W81XWH-15-1-0272, W81XWH2010578, W81XWH-15-1-0271, W81XWH-20-10576), and a Cottrell Scholar SEED Award (B.M.G., RCSA). W. M. Snow acknowledges support from US National Science Foundation grant PHY-1914405 and the Indiana University Center for Spacetime Symmetries.

%%%%%%%%%%%%%%%%%%%%%%%%%%%%%%%%%%%%
\bibliography{ref2.bib}
\bibliographystyle{apsrev4-1}

\end{document}

\title{Two-Orders-of-Magnitude Improvement in the Total Spin Angular Momentum of $^{131}$\/Xe Nuclei Using Spin Exchange Optical Pumping}

\author{Michael J. Molway}
\affiliation{%
Department of Chemistry and Biochemistry, Southern Illinois University, Carbondale, IL 62901, USA
}%

\author{Liana Bales-Shaffer}
\affiliation{%
Department of Chemistry and Biochemistry, Southern Illinois University, Carbondale, IL 62901, USA
}%

\author{Kaili Ranta}
\altaffiliation[Current address: ]{School of Medicine, The Ohio State University, Columbus, OH, USA}
\affiliation{%
Department of Chemistry and Biochemistry, Southern Illinois University, Carbondale, IL 62901, USA
}%

\author{Dustin Basler}
\affiliation{%
Department of Chemistry and Biochemistry, Southern Illinois University, Carbondale, IL 62901, USA
}%

\author{Megan Murphy}
\affiliation{%
Department of Chemistry and Biochemistry, Southern Illinois University, Carbondale, IL 62901, USA
}%

\author{Bryce E. Kidd}
\affiliation{%
Department of Chemistry and Biochemistry, Southern Illinois University, Carbondale, IL 62901, USA
}%

\author{Abdulbasit Tobi Gafar}
\affiliation{%
Department of Chemistry and Biochemistry, Southern Illinois University, Carbondale, IL 62901, USA
}%

\author{Justin Porter}
\affiliation{%
Department of Chemistry and Biochemistry, Southern Illinois University, Carbondale, IL 62901, USA
}%

\author{Kierstyn Albin}
\affiliation{%
Department of Chemistry and Biochemistry, Southern Illinois University, Carbondale, IL 62901, USA
}%

\author{Boyd M. Goodson}
 \email{bgoodson@chem.siu.edu}
\affiliation{%
Department of Chemistry and Biochemistry, Southern Illinois University, Carbondale, IL 62901, USA
}%

\author{Eduard Y. Chekmenev}
\affiliation{
Department of Chemistry, Integrative Biosciences (Ibio), Karmanos Cancer Institute (KCI), Wayne State University, Detroit, Michigan 48202, USA, 
Russian Academy of Sciences, 
Leninskiy Prospekt 14, Moscow, 119991, Russia
}%

\author{Matthew S. Rosen}
\affiliation{
A.A. Martinos Center for Biomedical Imaging, Massachusetts General Hospital and Harvard Medical School, Boston, MA 02129; Department of Physics, Harvard University, Cambridge, MA 02138, USA
}%

\author{W. Michael Snow}
\affiliation{
Department of Physics, Indiana University, Bloomington, IN, USA
}%

\author{James Ball}
\affiliation{
School of Medicine, University of Nottingham, Queens Medical Centre, Nottingham, UK
}%

\author{Eleanor Sparling}
\affiliation{
School of Medicine, University of Nottingham, Queens Medical Centre, Nottingham, UK
}%

\author{Mia Prince}
\affiliation{
School of Medicine, University of Nottingham, Queens Medical Centre, Nottingham, UK
}%

\author{Daniel Cocking}
\affiliation{
School of Medicine, University of Nottingham, Queens Medical Centre, Nottingham, UK
}%

\author{Michael J. Barlow}
\affiliation{
School of Medicine, University of Nottingham, Queens Medical Centre, Nottingham, UK
}%

\date{\today}

%Valid PACS numbers may be entered using the \verb+\pacs{#1}+ command.

%82.56.-b  NMR in chemical physics
%32.80.Xx  optical pumping of atoms
%87.61.-c  magnetic resonance imaging in medical physics
%42.55.Px  diode lasers 

% may need to add nuclear physics PACS codes?
%\pacs{32.80.Xx, 82.56.-b, 87.61.-c, 42.55.Px}% PACS, the Physics and Astronomy
                             % Classification Scheme.

Supplemental Information (SI) Document for

\vspace{1.0cm}

\maketitle

\newpage

\section{Experimental Methods}

\subsection{SEOP Apparatus}
\begin{figure}[h]
\centering
    \includegraphics[width=0.8\textwidth]{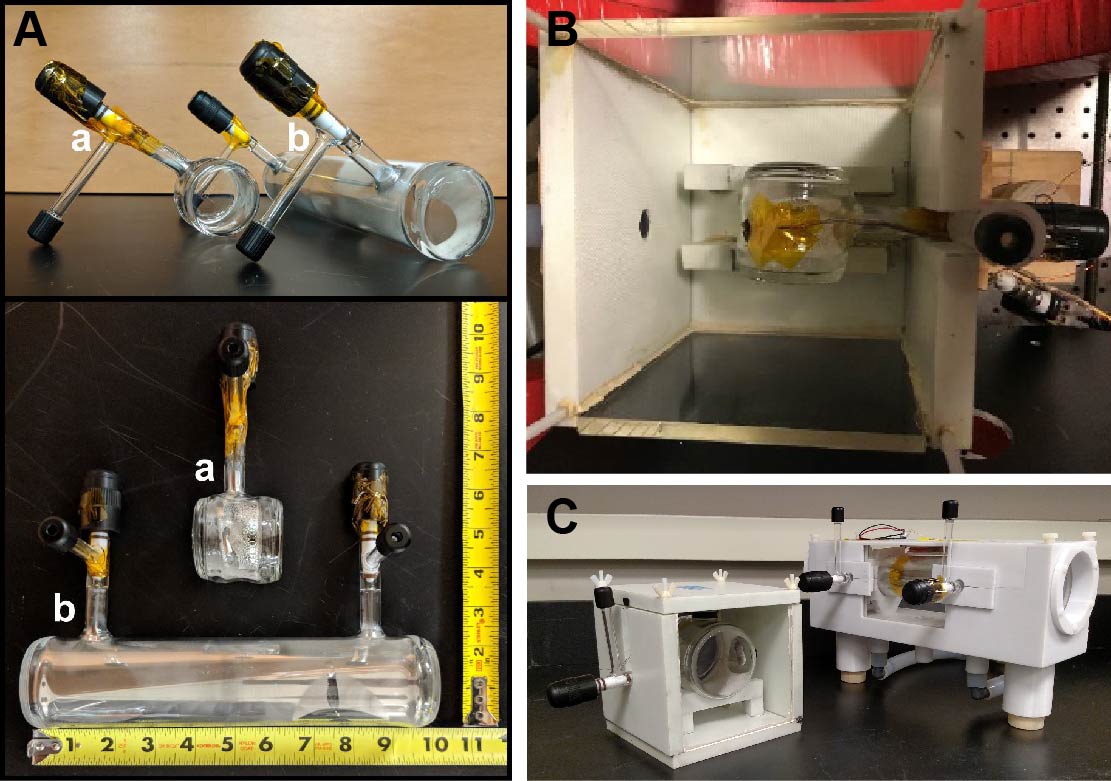}
    \caption{(A) Photos showing ``front'' (top) and ``top'' views of the 2" $\times$ 2" cell (a) used in this work; a 2" $\times$ 10" ``clinical-scale" (0.5 L) cell (b) typically used in our $^{129}$\/Xe hyperpolarization experiments is shown for comparison. (B) Photo of a 2" $\times$ 2" cell mounted in the Garolite oven.  (C) Photo showing assembled Garolite (left) and 3D-printed polycarbonate (right) forced-air ovens, containing a 3" $\times$ 3" and a 2" $\times$ 10" cell, respectively.  
    }
    \label{fig:cells_and_ovens}
\end{figure}

\begin{figure}[h]
\centering
    \includegraphics[width=0.8\textwidth]{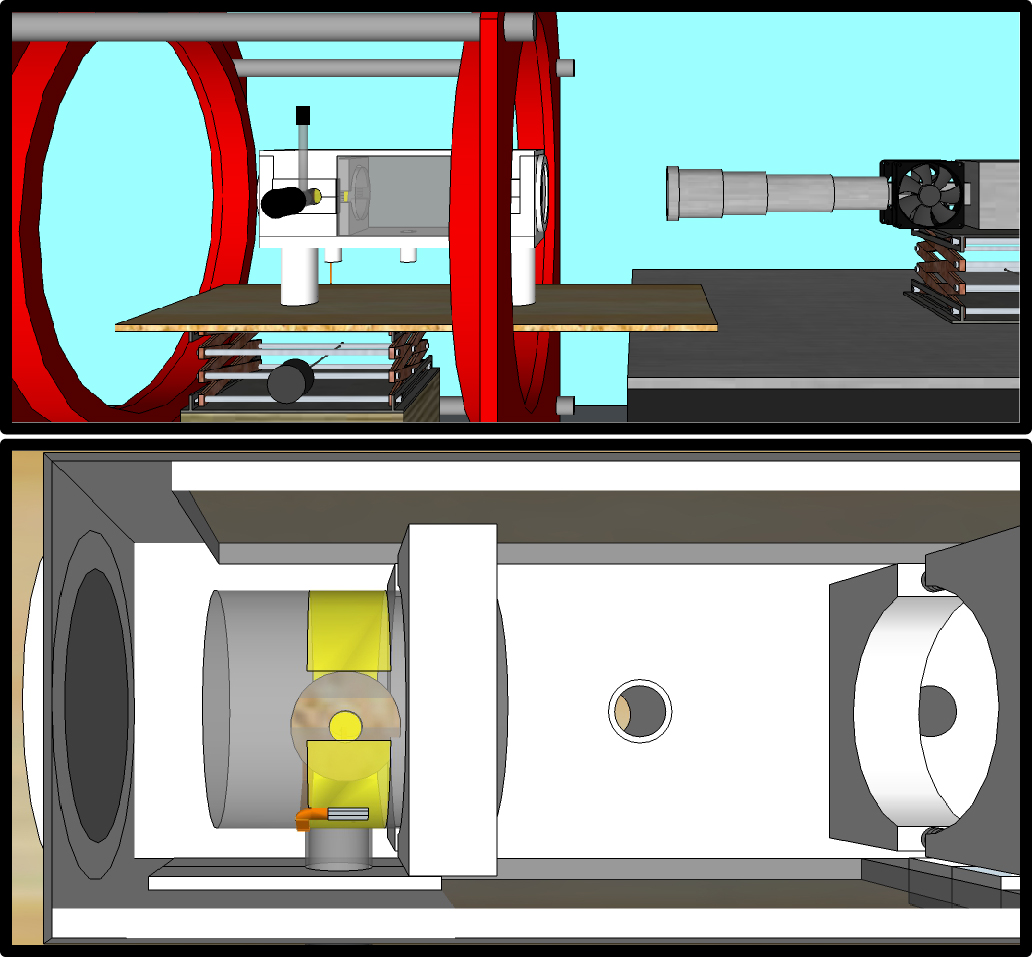}
    \caption{({\it Top}) CAD drawing of the SEOP apparatus, here with the 2" cell (gray cylinder) mounted in the polycarbonate 3D-printed forced air oven (white) and a QPC laser with mounted beam-expanding / polarizing optics (silver, right). The HC pair is shown in red (items not shown include the NMR spectrometer, the temperature controller, and PSUs). ({\it Bottom}) Top-view close-up of the interior of the 3D-printed oven, showing the 2" cell mounted near the rear, and thus in the most homogeneous region of the magnetic field.  A thermocouple (orange wire) is mounted above the stem with Kapton tape (yellow); the surface detection coil is shown as the circular object beneath the transparent cell.    A retro-reflecting mirror is mounted within the rear wall of the oven (dark gray, left).  }
    \label{fig:setup}
\end{figure}

All cells were custom-made from Pyrex by Midrivers glassblowing (Saint Peters, Missouri), with commercial stopcock assemblies (Chemglass) welded as side-arms to permit loading of contents; Fig. \ref{fig:cells_and_ovens}A).  Unless stated otherwise, all data were taken from uncoated cylindrical (2" $\times$ 2", i.e. 5.08 cm i.d. $\times$ 5.08 cm long) cells ($V$\/$\approx$\/103 cm$^3$).  Fig. \ref{fig:cells_and_ovens}A also shows a ``standard'' 10" long cell (500 cc, ``clinical scale'' \cite{NikolaouhighP, NikolaouhighP2}).  Cells were loaded with a small droplet (few hundred mg) of either Rb or Cs in an N$_2$\/-atmosphere glovebox. Then, a thin alkali metal film was then distributed with a heat-gun, being sure to keep the metal film off of the front and back glass windows. Cells were then loaded with gas (unless stated otherwise, 300 torr isotopically enriched $^{131}$\/Xe (84.4\%, Berry and Associates / Icon Isotopes) and 300 torr ultrapure N$_2$ (Airgas)). The cell was mounted in either a 3D-printed polycarbonate \cite{nikolaou_JACS_2014} or home-built Garolite forced-air oven (Fig. \ref{fig:cells_and_ovens}B,C); maximum temperatures used for the polycarbonate and Garolite ovens were 135 $^{\circ}$C and 225 $^{\circ}$C, respectively. Exterior cell wall temperature was monitored using a thermocouple fixed to the cell wall using Kapton tape, and the temperature measurement provided feedback for the oven's temperature controller (Berme; heat source is a Omega heat pipe, P/N: AHP-7561).  The oven was mounted such that the cell was centered within the homogeneous region of the magnetic field generated by a 22" Helmholtz coil (HC) pair (Walker Magnetics; Fig. \ref{fig:setup}) driven by an Agilent DC power supply.
$^{131}$\/Xe or $^{129}$\/Xe polarization was monitored {\it in situ} using a low-field NMR spectrometer (Magritek Aurora, nominal $^{131}$\/Xe NMR frequency: 20.9, 46, or 66 kHz, corresponding to magnetic fields of approximately 60, 130, or 190 G, respectively) connected to a 1" diameter surface detection RF coil mounted to the bottom of the cell (Fig. \ref{fig:setup}).  

To perform SEOP, resonant laser light was provided by one of three lasers used primarily in the present work tuned to the D1 line of Rb (794.8 nm) or Cs (894.3 nm).  Each laser design uses volume holographic gratings to spectrally narrow the laser output.  Before being directed into the cell, the light output of all three lasers was expanded and rendered circularly polarized using a polarizing beam-splitter cube and quarter-wave plate.  For Rb, the first laser used was a $\sim$\/150 W microchannel-cooled (MCC) laser (QPC Lasers) with spectral width (defined by the full width half maximum, FWHM) of 0.29 nm. When the Rb/$^{131}$\/Xe cell was first created, it was found that this laser deposited too much heat into the cell for stable SEOP above $\sim$\/90 $^{\circ}$\/C; thus, the laser's commercial optical train was fitted with a 3" expander and the widened beam was partially blocked with a 1.75" aperture, reducing the power on cell down to $\sim$\/100 W.  Later Rb/$^{131}$\/Xe SEOP experiments were performed with an Ultra-500 (U500) QPC laser with FWHM of 0.18 nm operated between 130-145 W (2" beam diameter). For all Cs/$^{131}$\/Xe experiments reported here, an OptiGrate ultra-narrow laser with FWHM of $\sim$\/50 pm was used (power on cell was $\sim$\/50 W).  

For some experiments, the spectral profile of near-IR light transmitted through the cell was monitored via fiber optic mounted behind a retro-reflecting mirror; the fiber optic was connected to an Ocean Optics near-IR spectrometer (HR2000 or HR4000, $\sim$\/0.04 nm resolution). Details concerning calculation of Xe polarization are provided below. Please see Refs. \cite{whitingJMR,saha} for other details concerning the SEOP apparatus. 

\subsection{Measurement of $^{131}$\/Xe Polarization Dynamics}
For a given SEOP cell (Rb or Cs), low-field {\it in situ} $^{131}$\/Xe NMR signals were collected during SEOP by first re-zeroing the $^{131}$\/Xe magnetization by saturating the signal using a ``crusher" sequence comprising a train of $\ge$\/100 long RF pulses (e.g. 9 ms) with quick repetition times (e.g. 0.2 s). Immediately following saturation, a $^{131}$\/Xe NMR spectrum was manually collected after a programmed delay time to obtain a single-shot spectrum. For quantification of polarization dynamics, such spectra were averaged either 6 or 24 times by repeating the above process for a given programmed delay time, which was systematically varied between 1 and 64 s.  Before fitting the data, a delay of 1.75 s (the average latency period measured for manual switching between ``crusher" and single-acquisition pulse sequences) was added to the programmed delay time.  This process was repeated at each temperature (once the setup had reached steady state) of 75, 85, 95, 105 and 115 $^{\circ}$\/C (although temperatures above 105 $^{\circ}$C were not stable for Rb SEOP because of ``Rb runaway" effects \cite{witte,zook,nikolaou_maps_2014}, reflecting in part the greater total laser flux and concomitant cell heating in those experiments). Averaged NMR signals from these scans were then used to calculate integrals and then fit to the equation: 

\begin{equation}
I(t)=I_{\infty} [1-\mathrm{exp}(-t\Gamma_{SEOP})] ,
    \label{eq:poldyneq}
\end{equation}

\noindent where $\Gamma_{SEOP}$ (the $^{131}$\/Xe polarization build-up rate constant) and $I_{\infty}$ (the steady-state integral of the $^{131}$\/Xe signal, $\int_{\infty}(^{131}\mathrm{Xe})$\/, were free-floating fit parameters.  Values for $\Gamma_{SEOP}$ and $\int_{\infty}(^{131}\mathrm{Xe})$\/ may be used to extract values for parameters governing the $^{131}$\/Xe polarization dynamics and the absolute polarization, as discussed in greater detail below in SI Sections 2 and 3, respectively.

\newpage

\section{Estimating Contributions to $^{131}$\/Xe Spin Relaxation and Spin Exchange}
Approaches that are described in the main document to estimate the contributions from $^{131}$\/Xe spin relaxation and alkali metal (AM)/$^{131}$\/Xe spin exchange to rate constants extracted from $^{131}$\/Xe polarization dynamics data are discussed in greater detail below. 

\subsection{Developing a simple model of $\Gamma_{Xe}$ for $^{131}$\/Xe}
\subsubsection{Revisiting gas-phase $^{129}$\/Xe spin relaxation.}
Before exploring contributions to gas-phase $^{131}$\/Xe nuclear spin relaxation, it is useful to consider the corresponding contributions to relaxation for gas-phase $^{129}$\/Xe.  Extensive discussions of nuclear spin relaxation mechanisms for gas-phase $^{129}$\/Xe and $^{131}$\/Xe can  be found in Refs. \cite{Brinkmann1962,adrian1965,Volk1980,Kwon1981,Raftery1994,Butscher1994,Chann2002,Anger2008,Stupic2011,Herman1965}, and only the most pertinent aspects will be mentioned here.  Briefly, we begin with a simplified version of Anger et al.'s formulation of the $^{129}$\/Xe nuclear spin auto-relaxation rate  \cite{Anger2008}:

\begin{equation}
    \Gamma_{Xe} =  \Gamma_t + \Gamma_p + \Gamma_w ,
    \label{eq:129Xerelax}
\end{equation}
%\vspace{0.5cm}

\noindent where $\Gamma_{Xe} = (T_1)^{-1}$\/, and $\Gamma_t$\/ and $\Gamma_p$ are the intrinsic relaxation contributions from transient Xe-Xe collisions (the binary term) and persistent Xe-Xe dimers (the three-body term), respectively; $\Gamma_w$ is the term arising from wall collisions. This formulation does not consider other potential contributions (e.g. from diffusion through field gradients or collisions with other non-Xe gas-phase species), all of which will generally be negligible under conditions relevant here.  A semi-empirical variant of 
Eq. \ref{eq:129Xerelax} provided in Ref. \cite{Anger2008}, which expands the intrinsic components of $^{129}$Xe relaxation, can be written as: 
\begin{equation}
    \Gamma_{Xe}(129) =  \alpha \mathrm{[Xe]} + \beta [1+\delta B_0^2] \left( 1+r\frac{\mathrm{[B]}}{\mathrm{[Xe]}} \right)^{-1} + \Gamma_w ,
    \label{eq:129Xerelax2}
\end{equation}
%\vspace{0.5cm}

\noindent where: $\alpha$ is the rate constant for relaxation from transient Xe dimers (explicitly reflecting the linear dependence of $\Gamma_t$ on [Xe], the xenon density in amagats), and for $^{129}$\/Xe is equal to (56.1 h)$^{-1}$\/; $\beta$ is the rate constant for relaxation from persistent Xe dimers, and for $^{129}$\/Xe is equal to (4.59 h)$^{-1}$\/; $\delta$ is a small coefficient that modulates a weak quadratic dependence of the persistent dimer term on magnetic field, $B_0$\/, and is equal to 3.65$\times$\/10$^{-3}$; [B] is the density of a given second gas in amagats (assuming a binary mixture); and $r$ is a ratio quantifying the relative efficiency with which the second gas (B) breaks up persistent Xe-Xe dimers compared to that of Xe itself; for binary mixtures where B=N$_2$\/, $r$\/=0.51 \cite{Anger2008}. 

For $^{129}$\/Xe ($I=1/2$) the intrinsic terms are known to be dominated by the spin-rotation interaction, with an additional contribution to the persistent-dimer term coming from the chemical-shift-anisotropy (CSA) interaction---hence the quadratic field dependence in Eq. \ref{eq:129Xerelax2}.  Additionally, whereas $\Gamma_t$ depends on [Xe], $\Gamma_p$ is independent of [Xe] and instead depends on the fraction of Xe versus other gases in the mixture. While overall the temperature dependence (originating from the modulation of Xe dimer formation/break-up rates) of the intrinsic terms was expected to be weak ($\sim T^{-1/2}$\/), a poorly understood $\sim T^{-2}$\/ dependence  was experimentally observed for $\Gamma_p$ \cite{Anger2008}.  $\Gamma_w$\/, on the other hand, arises from dipolar interactions between $^{129}$\/Xe and other spins (e.g. paramagnetic centers) on the inner surfaces of the container and is not expected to have a dependence on [Xe] (particularly if the average time for Xe to reach the container walls is short compared to $T_1$\/), but should be linearly proportional to the container's surface-to-volume ratio (S/V) and have an exponential (Arrhenius) dependence on temperature. 

\subsubsection{Applying the approach to gas-phase $^{131}$\/Xe spin relaxation.}
Unlike $^{129}$\/Xe, spin relaxation of $^{131}$\/Xe ($I=3/2$) is dominated by interactions between the nucleus's strong electric quadrupole moment and local electric field gradients. Nevertheless, the fluctuations in these field gradients (as felt by $^{131}$\/Xe) are modulated by the same types of collisions as those affecting $^{129}$\/Xe relaxation, giving rise to the same overall physical contributions to spin relaxation shown in Eq. \ref{eq:129Xerelax}---albeit with far greater overall rates.  For example, Brinkmann et al. reported a contribution from transient binary collisions, $\Gamma_t = \alpha$\/[Xe], with $\alpha$\/=3.96$\times10^{-2}$\/ s$^{-1}$\/, corresponding to a theoretical $^{131}$\/Xe $T_1$\/---even if it were the only contribution---of only 68 s at our room-temperature loading pressure of 300 torr Xe (0.37 amagat), compared to many hours for $^{129}$\/Xe.  The temperature dependence of this term is expected to be weak \cite{adrian1965}, and for simplicity will be ignored here.  However, contributions from persistent Xe dimers and wall collisions may be relatively large \cite{Stupic2011}, and must also be considered (as with $^{129}$\/Xe). Correspondingly, here we re-write Eqs. \ref{eq:129Xerelax} and \ref{eq:129Xerelax2} as:

\begin{equation}
    \Gamma_{Xe}(131) =  \alpha \mathrm{[Xe]} + \beta(B_0, T) \chi_{\mathrm{Xe}}'   + \Gamma_w ,
    \label{eq:131Xerelax}
\end{equation}
%\vspace{0.5cm}

\noindent where the term in parenthesis in Eq. \ref{eq:129Xerelax2} has been re-arranged to be represented as $\chi_{Xe}'$, effectively a ``reduced mole fraction" of Xe of the form: 

\begin{equation}
\chi_{\mathrm{Xe}}' = \frac{\mathrm{[Xe]}}{\mathrm{[Xe]}+r_B[\mathrm{B}]+r_A[\mathrm{A}]+...}, 
    \label{eq:chiXe}
\end{equation}
%\vspace{0.5cm}
\noindent which in our case of binary gas mixtures simplifies to 

\begin{equation}
\chi_{\mathrm{Xe}}' = \frac{\mathrm{[Xe]}}{\mathrm{[Xe]}+0.51[\mathrm{N}_2]},
    \label{eq:chiXe2}
\end{equation}
%\vspace{0.5cm}
\noindent if we assume that the $r$ value (quantifying the relative break-up efficiency of N$_2$ for persistent xenon dimers) should be the same for $^{131}$\/Xe as $^{129}$\/Xe.  In Eq. \ref{eq:131Xerelax}, we also assume that $\Gamma_w$\/(131) will be independent of gas mixture (as it is for $^{129}$\/Xe \cite{Anger2008}) but will have the same qualitative dependence on $T$ and $S/V$ as $\Gamma_w$\/(129) \cite{Butscher1994, Stupic2011} (see further below). Accordingly, we also expect that for $\Gamma_p$\/, $\beta$ in Eq. \ref{eq:131Xerelax} will likely have a dependence on $T$ and $B_0$\/ (but will be independent of [Xe] \cite{Stupic2011})---such that faster relaxation contributions will likely be observed at higher fields and lower temperatures.  

With the above estimate for $\Gamma_t$ in hand for our conditions, we thus also need estimates for $\Gamma_p$ and $\Gamma_w$\/ in order to estimate overall $\Gamma_{Xe}$\/(131) contributions to our observed $\Gamma_{SEOP}$\/ polarization build-up rate constants.  Toward this end, importantly Stupic et al. (Meersmann and co-workers) \cite{Stupic2011} provided a number of careful $\Gamma_{Xe}$\/(131) measurements with fixed total pressure but varying Xe mole fraction. Given fixed values for $T$\/=290 K and $B_0$\/=9.4 T, subtraction of $\Gamma_t$ calculated for each mixture allows the residual relaxation rate to be plotted as a function of $\chi_{Xe}'$\/. A successful linear fit (according to Eq. \ref{eq:131Xerelax}) should then give estimates of $\beta$ and $\Gamma_w$\/ from the slope and intercept, respectively.  

\begin{figure}[h]
    \centering
    \includegraphics[width=0.8\textwidth]{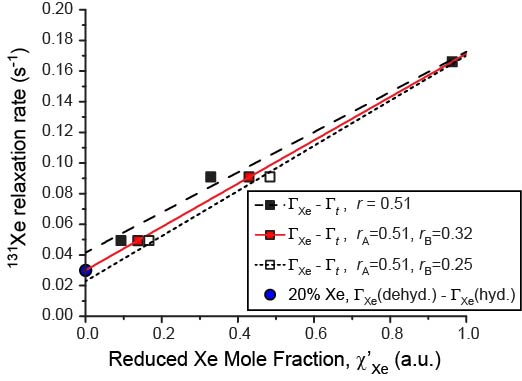}
    \caption{Experimental $^{131}$\/Xe spin-lattice relaxation rates ($\Gamma_{Xe}$\/=$1/T_1$\/) provided by Stupic et al. \cite{Stupic2011}, plotted as a function of $\chi_{Xe}'$\/, following subtraction of $\Gamma_t$\/ calculated for each Xe density. Reported conditions: $T$\/=290 K; $B_0$\/=9.4 T; Total pressure: 140 kPa; container: uncoated Pyrex tube (i.d.=12.6 mm); Xe:N$_2$:He gas mixtures (by fraction): 93:7:0; 20:5:75; 5:5:90.  Values for $\chi_{Xe}'$ for the three data points are calculated (and plotted) three different ways: assuming a single value for $r$\/=0.51 for all non-Xe components (black squares); assuming $r_A$\/=0.51 for N$_2$ \cite{Anger2008} and r$_B$\/=0.25 \cite{Chann2002} (white squares); and assuming  $r_A$\/=0.51 for N$_2$ and r$_B$\/=0.32 (red squares). A fourth point (blue circle) is provided by the relaxation rate measured from a 20:5:75 sample where the container walls had been first hydrated with water vapor to greatly reduce Xe surface adhesion; the value plotted is the experimental  $^{131}$\/Xe relaxation rate from that sample subtracted from the value measured from an identical gas mixture in a dehydrated (uncoated) sample tube. Because this difference (0.0296 $s^{-1}$\/) should remove contributions from both ([Xe]-dependent) $\Gamma_t$  and ($\chi_{Xe}'$\/-dependent) $\Gamma_p$ terms, it is plotted along the $y$\/-axis, and is thus treated as the effective lower-limit for $\Gamma_w$\/ for this sample. The value for r$_B$\/ was then optimized to find the smallest r$_B$ value that also resulted in a fit-line $y$\/-intercept of at least 0.0296 $s^{-1}$\/, yielding $r_B$\/=0.32.  Lines are corresponding linear fits. Fit results: $r$\/=0.51 (dashed black line): $\beta$\/=0.130, $\Gamma_w$\/=0.0417, $R^2$\/=0.99062; 
    $r_A$\/=0.51, $r_B$\/=0.25 (dotted black line): $\beta$\/=0.146, $\Gamma_w$\/=0.0231, $R^2$\/=0.99791;
    $r_A$\/=0.51, $r_B$\/=0.32 (red line): $\beta$\/=0.140, $\Gamma_w$\/=0.0302, $R^2$\/=0.99997.
    }
 \label{figure:meerrelaxdata}
\end{figure}

Three such plots (and corresponding linear fits) are provided below in SI Fig. \ref{figure:meerrelaxdata}.  In the first fit, we simplistically use a single value of $r=$\/0.51 (as if the all other gas atoms in the samples are N$_2$\/); this fit gives the result of $\beta$\/=0.130, $\Gamma_w$\/=0.0417 s$^{-1}$\/.  However, all but one of the samples in Ref. \cite{Stupic2011} included He as a third gas, and a lower $r$ value for He is expected. Indeed, an $r$ value of 0.25$\pm$0.08 was reported by Chann et al. in the context of $^{129}$\/Xe relaxation \cite{Chann2002}.  Plugging in this $r$ value for those samples containing He gas improved the fit, giving $\beta$\/=0.146, $\Gamma_w$\/=0.0231 s$^{-1}$\/.  As a cross-check of this fit, we can also consider a fourth measurement by Stupic et al. obtained from the same Xe fraction as their second point (20\% Xe), but loaded after additional treatment of the sample tube's glass surface with water vapor.  Hydration of the glass surface was expected to greatly reduce the surface adhesion for xenon atoms---and hence the contribution from $\Gamma_w$\/.  Subtraction of the  $^{131}$\/Xe relaxation rate measured from this sample with the hydrated surface, from that with the untreated glass surface, should remove contributions from both ([Xe]-dependent) $\Gamma_t$  and ($\chi_{Xe}'$\/-dependent) $\Gamma_p$ terms, and thus should effectively provide a lower limit for $\Gamma_w$\/ for this sample tube.  This value, 0.0296 $s^{-1}$\/, suggests that the fit result obtained with $r$(He)=0.25 underestimates $\Gamma_w$\/.  Thus, the $r$ value for He was then manually optimized to find the lowest value that also yielded a fit-line $y$\/-intercept of at least 0.0296 $s^{-1}$\/, thereby taking the information provided by the fourth data point into account. This outcome was achieved with $r$\/(He)=0.32, giving $\beta$\/=0.140, $\Gamma_w$\/=0.0302 $s^{-1}$\/.  We take these values moving forward, because doing so yielded the highest quality linear fit and also utilized all of the data points provided (this $r$\/(He) value is also within the error bars reported for the value in Ref. \cite{Chann2002}).

\vspace{1.0cm}

The next effort is to extrapolate the above values for $\beta$\/ and $\Gamma_w$\/ to our conditions.  If instead we simply estimate our persistent-dimer contribution using $\Gamma_p$\/=$\beta\cdot\chi_{Xe}'$\/ and the present value for $\beta$\/, then we would get $\Gamma_p$\/=0.0927 s$^{-1}$\/---an unrealistically large value that is significantly greater than $\Gamma_{SEOP}$\/ rates experimentally measured here.  However, this overestimate is not surprising, given that our measurements were obtained at higher temperature and much lower magnetic field.  Importantly, Stupic et al. also reported a $\Gamma_{Xe}$\/(131) value of 6.8$\times$\/10$^{-2}$\/ s$^{-1}$\/ measured at low field (millitesla regime) and high temperature (453 K) during SEOP in a larger Pyrex cell for their 93:7:0 Xe gas mixture. Subtraction of the binary contribution gives 3.5$\times$\/10$^{-2}$\/ s$^{-1}$\/ for the remaining contributions ($\Gamma_p+\Gamma_w$\/). 

In order to estimate the value for the $\Gamma_p$ contribution under these conditions (which are similar to those utilized for $^{131}$\/Xe SEOP here), we next consider the behavior of $\Gamma_w$\/ reported in Ref. \cite{Butscher1994}.  In that work, the experimentally measured contribution to $^{131}$\/Xe surface relaxation in a glass cell is modeled to have an Arrhenius dependence on temperature, governed by a xenon desorption activation energy of 0.12 eV.  From that temperature dependence \cite{Butscher1994}, a contribution to $\Gamma_w$\/ measured at 453 K would be reduced to $\sim$\/1.8\% of the value measured at 290 K.  Moreover, the reduction in S/V in comparing the containers used by Stupic et al. at 290 K versus 453 K should further reduce the $\Gamma_w$\/ value in the SEOP cell by another factor of $\approx$\/2.  Thus, an essentially negligible contribution of $\Gamma_w$\/=2.7$\times$\/10$^{-4}$ s$^{-1}$ can be estimated for their low-field SEOP cell measurement, yielding $\Gamma_p$\/=3.47$\times$\/10$^{-2}$ s$^{-1}$\/ for the same conditions.  

\begin{figure}[h]
    \centering
    \includegraphics[width=0.8\textwidth]{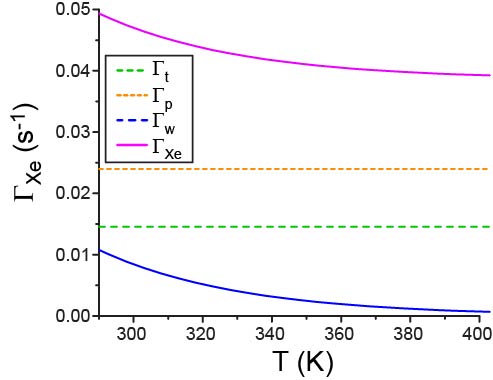}
    \caption{Estimated values for the different contributions to the $^{131}$\/Xe spin-lattice relaxation rate under our conditions. All three contributions ($\Gamma_t$\/, $\Gamma_p$\/, and $\Gamma_w$\/) are summed to give the overall estimated relaxation rate ($\Gamma_{Xe}$\/(131)) and its dependence on temperature. These curves are reproduced in Fig. 2(a) in the main document.  
    }
 \label{figure:relaxcurves}
\end{figure}

With an estimate of $\Gamma_p$ for the two different conditions (but the same Xe mole fraction) now in hand, $\beta$ for the low-$B_0$\/, high-$T$ conditions can now be estimated by taking the ratio of the two $\Gamma_p$\/ values (0.257), and multiplying by the known high-$B_0$, high-$T$ $\beta$ value above, to yield an estimate for $\beta$=0.036 under our similar conditions. The value for the $\Gamma_p$\/ term for our gas mixture and conditions can now be estimated for our value of $\chi_{Xe}'$ (0.662) using $\beta\cdot\chi_{Xe}'$\/=2.4$\times$\/10$^{-2}$ s$^{-1}$\/.  Finally, to estimate the wall contribution  over our temperature range, we take the 290 K value for $\Gamma_w$ estimated above from the data of Stupic et al. (reduced by $\approx$\/2.8-fold because of the smaller S/V ratio in our cells) and the exponential temperature dependence from Ref. \cite{Butscher1994}. Taken together, this gives our estimate for the temperature dependence for  $\Gamma_{Xe}$\/(131) under our conditions, as shown in SI Fig. \ref{figure:relaxcurves} (re-plotted in Fig. 2A of the main document).  Importantly, this estimate appears to be in reasonable agreement with the $\Gamma_{SEOP}$ build-up rate constants reported here, in that the  $\Gamma_{Xe}$\/ values approach (but are less than) those of $\Gamma_{SEOP}$, and $\Gamma_{SEOP}$ would be expected to be dominated by $\Gamma_{Xe}$\/ at the lowest temperatures.  Lastly, we note that this approach ignores any additional temperature dependence of $\Gamma_p$ over our temperature range (instead using the extrapolated value from Stupic et al. obtained at 453 K), and thus the calculated values for $\Gamma_{Xe}$\/ reported here are likely slight underestimates---particularly at lower temperatures.

\vspace{1.0cm}

% \vspace{5.0cm}

\subsection{Extracting Spin-Exchange Contributions from $^{131}$\/Xe Polarization Build-Up Rates}
Detailed experimental and theoretical studies of spin exchange between alkali metal vapors and xenon (both $^{129}$\/Xe and $^{131}$\/Xe isotopes) can  be found elsewhere (e.g., Refs. \cite{Stupic2011,Herman1965,Volk1980,Kwon1981,WalkerHapper,jau3papers,hughes2005,luo1999}), 
and key aspects will be described only briefly here. Whereas polarization of the alkali metal vapor's electron spins is treated here as being effectively instantaneous, the polarization of the Xe nuclear spins occurs much more slowly, and can be easily observed in real time via {\it in situ} low-field NMR. The NMR signal at a given time, $S(t)$, will be proportional to the $^{131}$\/Xe nuclear spin polarization, $P(t)$\/.  Under stable conditions, the Xe nuclear spin polarization will build up over time according to a simple exponential relationship:   

\begin{equation}
P_{Xe}(t) = P_{Xe}^{\infty}\cdot \left[1-\mathrm{exp}\left( -\Gamma_{SEOP}\cdot t\right) \right],
    \label{eq:Pbuildup}
\end{equation}
%\vspace{0.5cm}
\noindent where $\Gamma_{SEOP}$ is the experimentally derived hyperpolarization build-up rate constant for a given set of conditions and $P_{Xe}^{\infty}$ is the steady-state xenon polarization, given by:

\begin{equation}
P_{Xe}^{\infty} = \langle P_{AM}\rangle \cdot \frac{\gamma_{SE}}{\Gamma_{SEOP}},
    \label{eq:PXeinf}
\end{equation}
%\vspace{0.5cm}
\noindent where $\langle P_{AM}\rangle$ is the spatiotemporal average of the alkali metal vapor's electron spin polarization.

As with $^{129}$\/Xe, the experimentally derived $\Gamma_{SEOP}$ value for $^{131}$\/Xe under some given set of conditions (e.g. cell temperature, gas pressures, etc.) is given by: 
\begin{equation}
    \Gamma_{SEOP} =  \Gamma_{Xe}(131) + \gamma_{SE},
    \label{eq:gamma_buildup}
\end{equation}
%\vspace{0.5cm}
\noindent where $\gamma_{SE}$ is the spin-exchange rate between the alkali metal (here Rb or Cs) and $^{131}$\/Xe. Thus, the independent knowledge of the 
$^{131}$\/Xe nuclear spin auto-relaxation rate (provided above) can be used to extract $\gamma_{SE}$ values from experimental polarization-build-up rates. In turn, $\gamma_{SE}$ is given by:
\begin{equation}
    \gamma_{SE} = \mathrm{[AM]} \cdot \gamma ^{\prime} ,
    \label{eq:gamma_SE}
\end{equation}
%\vspace{0.5cm}
\noindent where $\gamma^{\prime}$ is the per-atom spin-exchange rate (effectively, the total spin-exchange cross-section, which is the sum of $\langle \sigma \overline{v} \rangle$\/, the velocity-averaged binary spin-exchange cross section, and a ``three-body" term arising from transient van der Waals AM-Xe complexes). 

Figure 2(a) in the main document shows the values for $\Gamma_{SEOP}$ (derived from exponential fits of the $^{131}$Xe polarization build-up curves in Fig. 1 of the main document), plotted along with the various estimated contributions to $\Gamma_{Xe}(131)$\/ shown above.   As mentioned above, $\Gamma_{SEOP}$ is dominated by the estimates for $\Gamma_{Xe}(131)$\/---particularly at lower temperatures---in good agreement with expectations.  One exception is the anomalous first point in the Rb/Xe data.  While higher $\Gamma_{Xe}(131)$\/ rates (and hence, total $\Gamma_{SEOP}$\/ values \cite{Volk1980,Butscher1994}) are indeed expected at lower temperatures, neither the point's large absolute value nor its difference from the corresponding point in the Cs data set can be explained by the $\Gamma_{Xe}(131)$\/ temperature dependence.  While this dataset had the lowest SNR of all of the runs, the origin of the anomalous behavior is uncertain. One possible explanation could be an unexpectedly high AM density (e.g. possibly owing to direct laser heating of Rb on the inner cell wall during that earlier run).

As shown below, analysis of a set of $\Gamma_{SEOP}$ values can in principle provide an estimate of the AM/Xe spin-exchange cross-section.  Here, we are interested in three specific aspects: (1) the absolute magnitude of the SE cross-section for $^{131}$Xe (and its relative size compared to $^{129}$Xe), to determine the viability of using SEOP to prepare bulk hyperpolarized $^{131}$Xe for envisioned NOPTREX experiments or other applications; (2) to inform future modeling of $^{131}$Xe SEOP; and (3) to compare the cross-section values attained with a given alkali metal.  

As pointed out by Volk and co-workers \cite{Volk1980,Kwon1981}, the theory of Herman \cite{Herman1965} (which originally considered the calculation of spin-exchange cross-sections between electrons of Rb vapor atoms and the nuclei of other gases including $^3$\/He, $^{21}$\/Ne, $^{83}$\/Kr, H$_2$, and D$_2$\/)
may be used to predict the ratio of the spin-exchange cross sections for isotopes of the same element, say Xe (for spin-exchange with atoms of a given metal vapor), according to: 
\begin{equation}
\frac{\gamma^{\prime}_{I}}{\gamma^{\prime}_{K}} = \frac{I(I+1)\gamma_I^2}{K(K+1)\gamma_K^2},
    \label{eq:ratioofcross}
\end{equation}
%\vspace{0.5cm}
\noindent where $I$ and $K$ are the nuclear spins for the isotopes in question and $\gamma_{I(K)}$ are the corresponding gyromagnetic ratios---which for $^{129}$\/Xe ($I=1/2$\/) and $^{131}$\/Xe ($K=3/2$\/) predicts a ratio of $\approx$\/2.28.  Yet Volk and co-workers reported an experimental $\gamma^{\prime}$\/ value for Rb/$^{131}$\/Xe (in cells with only 0.5 torr of enriched $^{131}$\/Xe) that was $\sim$\/16-fold smaller than that measured (separately) for Rb/$^{129}$\/Xe---an initial result that was rationalized as possibly originating from a hypothesized additional (e.g. quadrupolar) interaction that reduces the efficiency of spin exchange for $^{131}$\/Xe \cite{Volk1980}.  If true, such a large difference, particularly when combined with the much faster autorelaxation rates for $^{131}$\/Xe, would dim the prospects for the viability of using SEOP for bulk preparation of hyperpolarized $^{131}$\/Xe (e.g. for the envisioned NOPTREX experiments). However, in experiments from the same group where measurements were performed on $^{129}$\/Xe and $^{131}$\/Xe in the same Rb-containing cell (and also with 0.5 torr Xe), comparison of transverse relaxation rates (under conditions where spin-exchange was expected to be the dominant contribution) gave a smaller ratio of $\approx$\/3.56 \cite{Kwon1981}---in better agreement with the theoretical prediction. Mehring and co-workers \cite{Butscher1994} later reported a Rb/$^{131}$\/Xe spin-exchange cross-section that was $\sim$\/2-fold larger than that from the earlier work \cite{Volk1980}---albeit at significantly higher gas densities (13 mbar and 150 mbar of Xe and N$_2$\/, respectively, though significantly less than the densities utilized in the present work).   

Following Eq. \ref{eq:gamma_buildup}, subtraction of $\Gamma_{Xe}(131)$ from the provided $\Gamma_{SEOP}$ values gives estimates for the $\gamma_{SE}$ values for both Rb and Cs data sets. In turn, these values can be plotted versus the alkali metal density [AM] for both Rb and Cs (see Fig. 2(b) in main document).  Given the absence of a direct experimental method for measuring [AM] for our apparatus, here [AM] values are estimated from temperature-dependent vapor-pressure curves from the literature \cite{SteckCsRb}. Stupic {\it et al.} reported a $\gamma_{SE}$ value for Rb/$^{131}$\/Xe of $\sim$\/1.7$\times$\/10$^{-2}$\/ s$^{-1}$\/ \cite{Stupic2011}, which is in the range of values shown in the figure---albeit at a higher temperature of 453 K (where, based on vapor pressure curves \cite{SteckCsRb}, it would be expected to have [Rb]$\approx$\/4$\times$\/10$^{14}$ atoms/cm$^{3}$\/); this data point was recorded for the 97:7:0 Xe:N$_2\/$:He gas mixture (150 kPa total pressure \cite{Stupic2011}).   In the case of $^{129}$\/Xe, greater values would be expected for spin-exchange rates at lower Xe partial pressures (and total pressures), because of the greater additive contribution of the three-body term.  Indeed, if SEOP spin-up rate constants from Stupic {\it et al.} are treated with the same analysis of $^{131}$\/Xe provided above, a somewhat larger $\gamma_{SE}$ rate would be estimated, at least for the value with the lowest Xe partial pressure.

We now provide additional details about the comparisons provided in the main document between the spin-exchange rates found in the present work and those in the literature.  It is known that actual alkali metal densities can vary systematically from those predicted from vapor pressure curves \cite{jau3papers,hughes2005,chann2002PRA}; thus, values extracted from such analyses (such as those presented here) should be regarded as only approximate. Nevertheless, to the degree that: (1) the alkali metal vapor pressure curves can at least provide reasonable estimates for the {\it changes} in [AM] over a given temperature range; (2) all of the contributions from $\Gamma_{Xe}(131)$\/ have been properly subtracted; and (3) any temperature dependencies of the two-body and three-body contributions to $\gamma_{SE}$ can be safely neglected over a given temperature range, then the slope of a linear fit of the $\gamma_{SE}$ data (forced through zero, since the spin-exchange rate should be zero when the alkali metal vapor density is zero) should provide an estimate of $\gamma^{\prime}$\/ for each alkali metal / $^{131}$\/Xe combination. Excluding the anomalous first point, the Rb data exhibit a steeper dependence on the estimated [AM] value compared to Cs under our conditions.  Fitting the remaining Rb $\gamma_{SE}$\/ points (and the origin) gives an estimate of $\gamma^{\prime}$\/(Rb/$^{131}$\/Xe)=5.1($\pm$0.4)$\times10^{-15}$\/ cm$^{3}\cdot$\/s$^{-1}$\/. The deviation from linearity observed in the Cs results at higher temperatures may reflect insufficient accounting for the temperature dependence of $\Gamma_{Xe}$\/(131)---particularly in the persistent-dimer term.  A steeper fall-off in $\Gamma_{Xe}$\/(131) would likely give rise to greater $\gamma_{SE}$ values at higher temperatures.
To obtain the most comparable fit for Cs---involving a similar range of [AM] and $\Delta T$\/ as that utilized here for $^{131}$\/Xe/Rb---the first three points (and the origin) were fit to give $\gamma^{\prime}$\/(Cs/$^{131}$\/Xe)=1.5($\pm$0.2)$\times10^{-15}$\/ cm$^{3}\cdot$\/s$^{-1}$.  These values can be compared with previous values from the literature:  Arguably the most straightforward comparisons can be made to our previous efforts to compare Rb/$^{129}$\/Xe and Cs/$^{129}$\/Xe spin exchange (given the rough similarities in Xe partial pressures, incident laser powers, cell volumes, and cell temperatures), where values of $\gamma^{\prime}$\/(Rb/$^{129}$\/Xe)=1.67($\pm$0.06)$\times10^{-15}$\/ cm$^{3}\cdot$\/s$^{-1}$ and $\gamma^{\prime}$\/(Cs/$^{129}$\/Xe)=2.6($\pm$0.1)$\times10^{-15}$\/ cm$^{3}\cdot$\/s$^{-1}$, respectively, were reported \cite{nick2011}. Considering the Cs results first, the present results---combined with the previous measurements---would yield a $\gamma^{\prime}$\/(129):$\gamma^{\prime}$\/(131) ratio of 1.7$\pm$\/0.3, in reasonable agreement with expectations based on the theory of Herman \cite{Herman1965}. Moreover, albeit under different conditions, Hughes and coworkers reported a range of values of $\gamma^{\prime}$\/(Cs/$^{129}$\/Xe)=(1.9-5.4)$\times10^{-15}$\/ cm$^{3}\cdot$\/s$^{-1}$ \cite{hughes2005}---which would again would be larger than the $\gamma^{\prime}$\/(Cs/$^{131}$\/Xe) value reported here by an amount that is in rough agreement with the theoretical prediction.  However, the value for $\gamma^{\prime}$\/(Rb/$^{131}$\/Xe) obtained above seems too high in two respects: first, it is significantly {\it larger} than both the corresponding $\gamma^{\prime}$\/(Rb/$^{129}$\/Xe) values reported in both Refs. \cite{nick2011} (mentioned above) and \cite{hughes2005} [$\gamma^{\prime}$\/(Rb/$^{129}$\/Xe)=(1.5-3.2)$\times10^{-15}$\/ cm$^{3}\cdot$\/s$^{-1}$]; second, in those works, a ~1.5-2-fold {\it smaller} $\gamma^{\prime}$ value was observed for Rb/$^{129}$\/Xe compared to Cs/$^{129}$\/Xe, and it is unclear why the trend would be reversed for $^{131}$\/Xe spin-exchange.  One possible explanation is that the differences in the nature of the laser excitation (and corresponding AM heating by the laser) used for Rb versus Cs in the present work gave rise to higher-than-expected [Rb] values, causing $\gamma^{\prime}$\/(Rb/$^{131}$\/Xe) to be over-estimated in the present work.  In any case, the values for the $^{131}$\/Xe spin-exchange cross-sections for both Cs and Rb appear to be high enough to readily enable the attainment of experimental AM/$^{131}$\/Xe spin-exchange rates that are sufficiently large to compete with the reported high $^{131}$\/Xe auto-relaxation rates---even at relatively high Xe densities.

\section{Calculating polarization of HP $^{131}$Xe}

\subsection{Introduction}

\begin{figure}[H]
    \centering
    \includegraphics[width=0.5\textwidth]{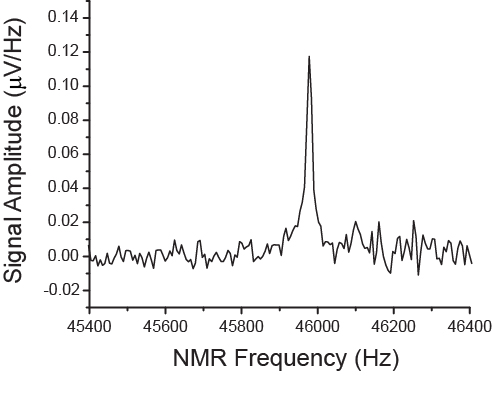}
    \caption{Example of a low-field $^1$\/H NMR spectrum (46 kHz) from a 10" cell containing thermally polarized water doped with CuSO$_4$ (60,000 scans; here shown with 5.83 Hz apodization).}
    \label{fig:waterref}
\end{figure}

Calculation of the $^{131}$Xe polarization level achieved for a given experiment ($P_{\mathrm{131Xe}}$) is performed using Eq. \ref{eq:master} below.  The polarization of the {\it in situ} low-field NMR signal from hyperpolarized (HP) $^{131}$Xe is determined by comparing its integral [$\int(^{131}\mathrm{Xe})$\/] with that of a sample of known polarization achieved at thermal equilibrium.  However, it is impractical to attempt detection of thermally polarized $^{131}$Xe gas because the NMR signal would be far too weak at the present magnetic fields. Instead, the HP signal is referenced to a signal from thermally polarized $^{1}$\/H spins in water, $\int(^{1}\mathrm{H})$\/, averaged over a period of time (see examples of $^1$\/H and $^{131}$\/Xe spectra in SI Figs. \ref{fig:waterref}, \ref{figure:etacell}, and \ref{fig:131andIRexamples}). This integral ratio can then be multiplied by the thermal $^1$\/H polarization, $P_{\mathrm{1H}}$\/, according to the relation:

\begin{equation}
    P_{\mathrm{131Xe}} = P_{\mathrm{1H}} \cdot
    \frac{\int\left(^{131}\mathrm{Xe} \right)}{\int\left(^{1}\mathrm{H}\right)} \cdot 
    \frac{[^{1}\mathrm{H}]}{[^{131}\mathrm{Xe}]}  \cdot \frac{\gamma_{\mathrm{H}}}{\gamma_{\mathrm{Xe}}}  \cdot \frac{I_{\mathrm{H}}}{I_{\mathrm{Xe}}}   \cdot C_{T2*} \cdot  \left(\frac{\eta_f \cdot \eta_{pulse}}{\eta_T \cdot \eta_{cell}}\right)  ,
\label{eq:master}
\end{equation}

\vspace{0.5cm}

\noindent which also contains several factors to correct for the differences in the nuclei, as well as several systematic contributions that depend on the nature of the experiment: $[^{1}\mathrm{H}]$ and $[^{131}\mathrm{Xe}]$ represent the spin concentrations of the xenon and hydrogen samples, whereas each $\gamma$ and $I$ value corresponds to the gyromagnetic ratio and spin of each nucleus, respectively.  Values used in our calculations are provided in Table \ref{tab:first_seven}. Additionally, $C_{T2*}$ is a correction factor determined by the full width half maximum (FWHM) values of the peaks in the HP and thermally polarized reference NMR spectra. This factor takes into account signal lost during the pre-acquisition delay that may result from differential dephasing rates of the two nuclei within that delay window (see, e.g., Ref. \cite{nikolaou_maps_2014}). The value of $C_{T2*}$ is typically close to unity, and is determined by Eq. \ref{eq:CT2*}:

\begin{equation}
    C_{T2*} = \mathrm{exp}\left(\frac{T_{AQ}}{T_{2}^{*}(\mathrm{Xe})}-\frac{T_{AQ}}{T_{2}^{*}(\mathrm{REF})}\right) ,
    \label{eq:CT2*}
\end{equation}

\vspace{0.5cm}
 
\noindent where $T_{AQ}$ is the acquisition delay and the $T_{2}^{*}$ values are the exponential decay constants for the time-domain free induction decay (FID) signals from the HP species (Xe) and reference $^1$\/H (REF) nuclei, respectively [$T_{2*}$ values are calculated using: $T_{2}^{*} = (\pi \cdot \mathrm{FWHM})^{-1}$]. Finally, the $\eta$ values represent systematic corrections that must be taken into account when different frequencies ($f$), sample temperatures ($T$), sample cell geometries ($cell$), and/or RF pulse tipping angles ($pulse$) are used for the HP and reference spectra.

\begin{table}[H]
    \centering
    \includegraphics[width=0.9\textwidth]{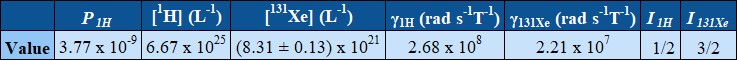}
    \caption{Values of the first seven terms and constants in Eq. \ref{eq:master} used to calculate $P_{131Xe}$ values for both Dataset(1) and Dataset(2). A partial pressure of 300 torr Xe (enriched to 84$\%$ $^{131}$\/Xe) was used for all data reported in this work. }
    \label{tab:first_seven}
\end{table}

We note that for each cell temperature, the integrals and uncertainties were obtained either by fitting the polarization dynamics data using the averaged integrals at each time point (thereby giving the $t=\infty$ values as the $^{131}$\/Xe integrals and fit error bars as the uncertainties for that temperature point) or, where necessary, by taking the values from single-point acquisitions (e.g. under conditions of Rb runaway).  All of the other terms for calculating polarization in Eqs. \ref{eq:master} and \ref{eq:CT2*} are summarized in Table \ref{tab:master}, displaying a single example each from Dataset(1) and Dataset(2).  The value for the integral of $^1$H is the same for both datasets as the same water reference signal is used. $\eta_{cell}$ is also the same as effectively identical cells were used in both experiments (more specifically, the same Rb cell was used in obtaining Dataset(2) as Dataset(1); the Cs cell used in obtaining Dataset(1) was nearly identical). $\eta_f$ is not needed for Dataset(2) as both the reference $^1$H signal and the $^{131}$Xe signal were acquired at the same frequency (46 kHz). $\eta_{pulse}$ is not needed for Dataset(2), because all spectra were acquired with 90$^\circ$ pulses; however, for Dataset(1), it was later determined that a 30$^\circ$ tipping angle was used, corresponding to a correction factor of $\eta_{pulse}=(\sin{30^{\circ}})^{-1}=2$.  Experiments performed to separately determine $\eta_{cell}$, $\eta_{pulse}$, $\eta_{T}$, and  $\eta_{f}$ are described below in this SI document. 

\begin{table}[H]
    \centering
    \includegraphics[width=\textwidth]{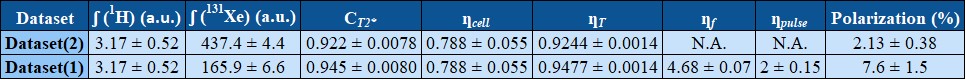}
    \caption{Examples of values from each Dataset used to calculate $P_{131Xe}$ values according to Eqs. \ref{eq:master} and \ref{eq:CT2*}. N.A. is reported for each term that was not used to calculate the polarization in Dataset(2) (value is effectively 1, because the same frequency (46 kHz) and pulse tipping angle (90$^\circ$) was used for both HP and reference spectra). }
    \label{tab:master}
\end{table}

\subsection{Pulse Tipping Angle Calibration}

\begin{figure}[h]
    \centering
    \includegraphics[width=0.8\textwidth]{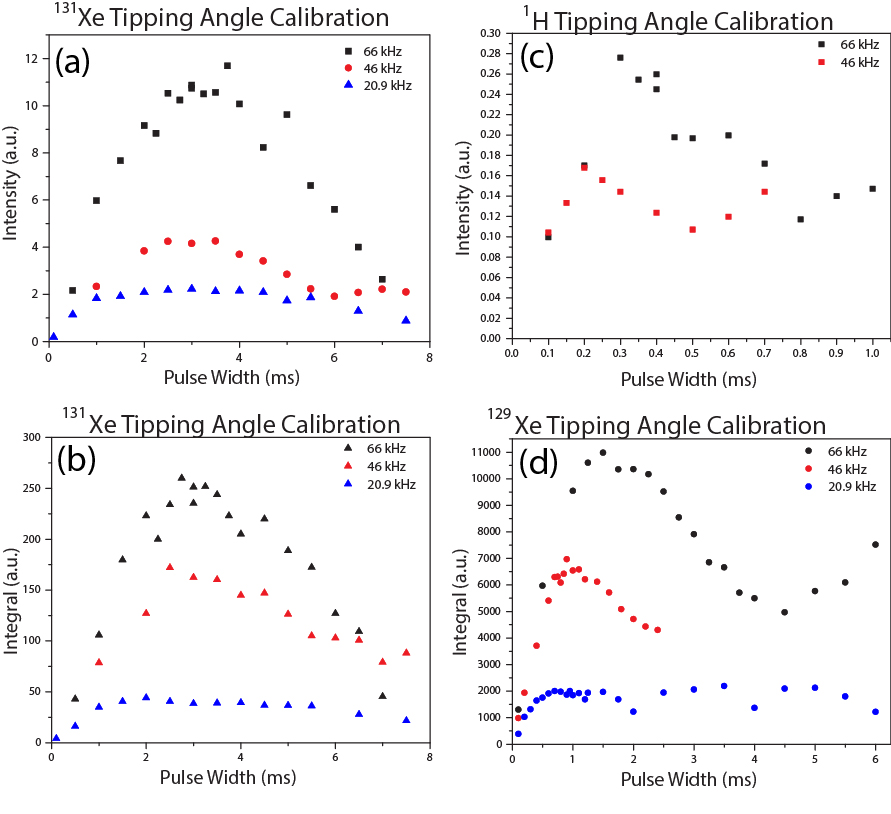}
    \caption{Selected RF pulse tipping-angle calibration curves for $^{131}$\/Xe (a,b), $^{1}$\/H (c), and $^{129}$\/Xe (d) at 20.9 (blue), 46 (red), and 66 kHz (black).  NMR Spectrometer: Magritek Aurora, with surface coil.  Pulse amplitude set at half the nominal value (5 V versus 10 V) to improve precision (linearity of the amplifier was determined separately).
    \label{fig:tipping}
}
\end{figure}

RF pulse tipping angles for the $^{131}$\/Xe, $^{129}$\/Xe, and $^{1}$\/H were measured at 46 kHz and 66 kHz; tipping angles were also calibrated for  $^{131}$\/Xe and $^{129}$\/Xe at 20.9 kHz (SI Fig. \ref{fig:tipping}). $^{1}$\/H tipping angles were calibrated using thermally polarized water (10" cell) with long-time signal averaging, whereas $^{131}$\/Xe and $^{129}$\/Xe tipping angles were measured using hyperpolarized xenon (single-scan acquisitions on 2" and 10" cells, respectively).   Values are summarized in Table \ref{fig:tip_angle_table}. 

 \begin{table}[H]
    \centering
    \includegraphics[width=0.7\textwidth]{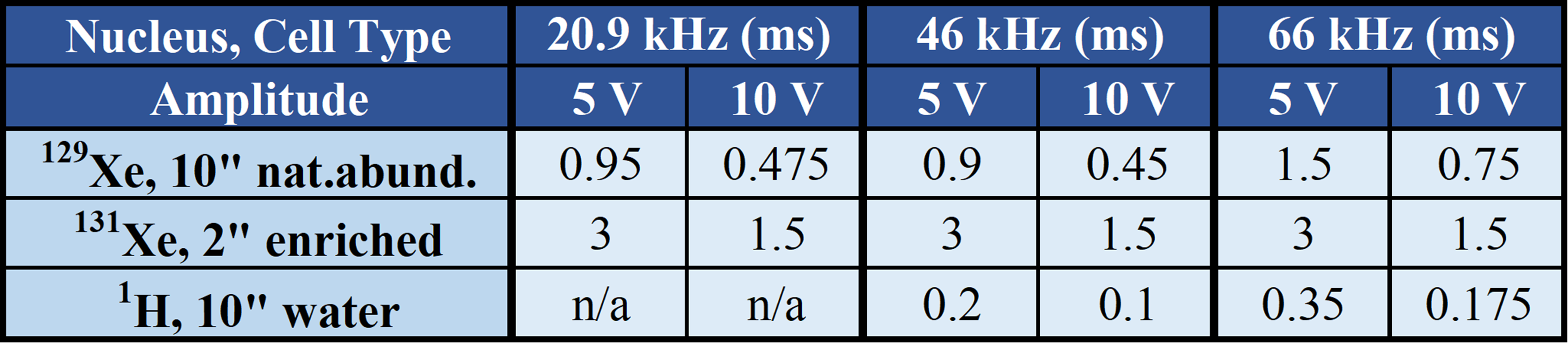}
    \caption{Summary of RF pulse widths corresponding to 90$^{\circ}$ tipping angles estimated from experimental data (e.g. curves in SI Fig. \ref{fig:tipping}) for 5 V pulse amplitudes for $^{131}$\/Xe, $^{129}$\/Xe, and $^{1}$\/H at the three relevant frequencies; 10 V values are the 5 V values divided by 2. Tabulated values were determined using both integral and peak intensity curves. Uncertainties were generally $\sim$\/8-17\%.  
    \label{fig:tip_angle_table}
}
\end{table}

\subsection{Determination of $\boldsymbol{\eta}_\mathbf{cell}$}

\begin{figure}[H]
\centering
    \includegraphics[width=\textwidth]{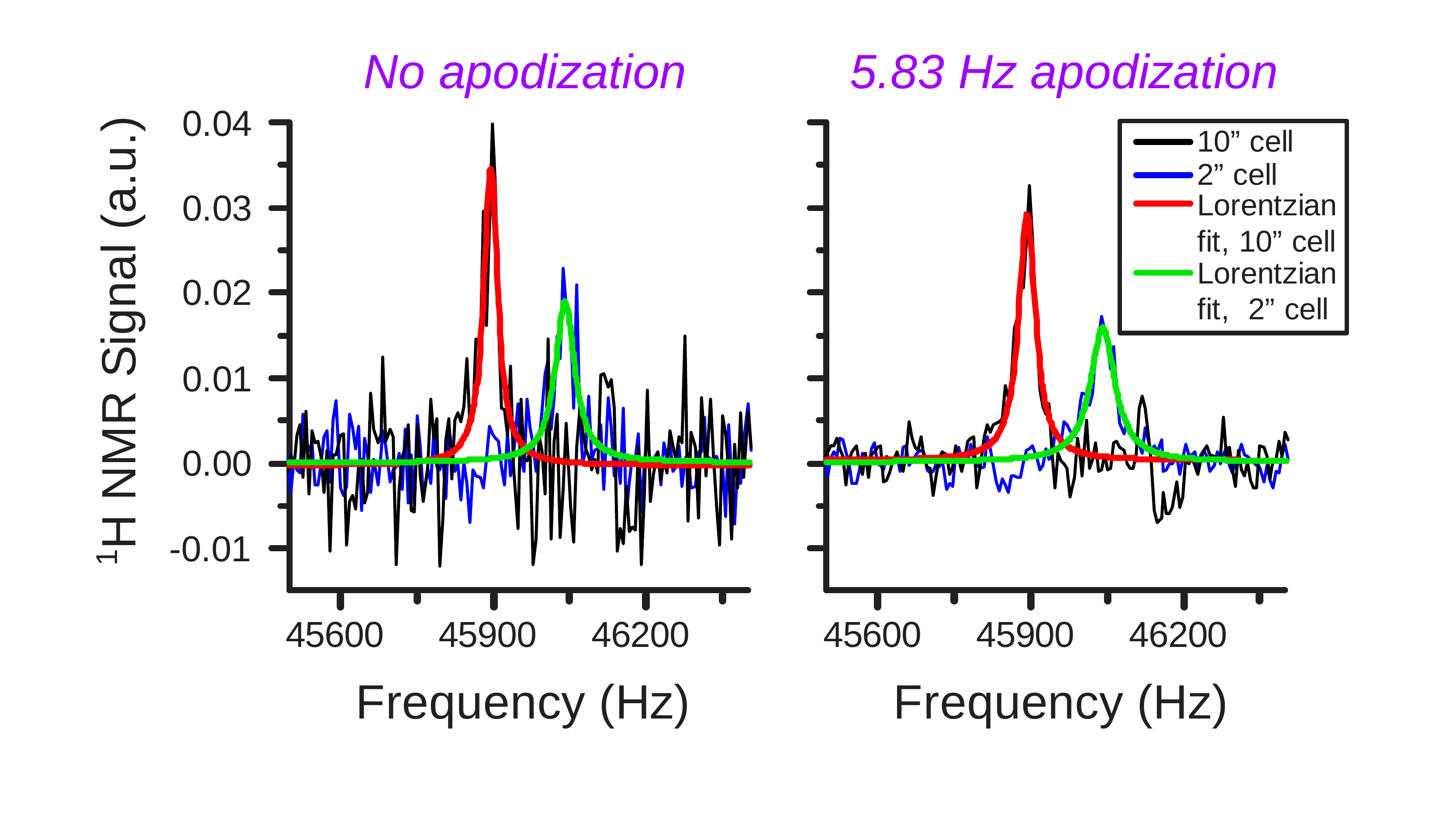}
    \caption{Low-field $^1$H NMR spectra from two different cylindrical cells (both 2" diameter, one 10" long (black), one 2" long (blue)) containing tap water; the 2"-long cell is the same cell as that used to obtain the HP $^{131}$\/Xe/Rb data, following removal of its contents.  The small frequency offset between the 10" cell spectrum (black) and the 2" cell spectrum (blue) reflects slight changes in the magnetic field strength over time.  Superimposed on the processed spectra are Lorenztian fits (red and green curves, respectively), performed after baseline correction. Left panel: spectra and fits without broadening. Right panel: same as left, but with 5.83 Hz exponential broadening, applied using Prospa software prior to export to Microcal Origin software for further processing / analysis. Acquisition parameters: pulse width: 0.1 ms; pulse amplitude: 10 V; bandwidth: 100 kHz; pre-acquisition delay: 3.8 ms; recycle delay: 0.6 s; number of scans: 12,000 and 18,000, respectively (Magritek Aurora spectra are auto-normalized by number of scans; note that a multi-second recycle delay to ensure complete re-thermalization of the polarization was not necessary because these spectra are not themselves used for absolute polarization calibration).  The ratio of the resulting integrals was used to calculate $\eta_{cell}$  = $\int (2") / \int (10")$ = 0.788 $\pm$ 0.055.}
    \label{figure:etacell}
\end{figure}

A $^{1}$\/H thermally polarized spectrum (e.g. that in SI Fig. \ref{fig:waterref} above) can be used to calibrate the absolute polarization of hyperpolarized $^{131}$\/Xe samples.  However, it first needs to be corrected for a systematic error potentially caused by using a cell with a different geometry (10" length) compared to the $^{131}$\/Xe cell (2" length; there are also small differences in diameter for these hand-blown cells).   The resulting difference in signal was anticipated to be small (because of the use of a small surface coil for pulse/detection), but still important to quantify; this correction factor is $\eta_{cell}$  in Eq. \ref{eq:master}.

For this purpose, the contents of the 2" cell (used for HP $^{131}$\/Xe) was cleaned out following the completion of SEOP experiments, and filled with tap water (instead of a CuSO$_4$ solution, to avoid the use of high concentrations of paramagnetic centers in a cell that could be reused for hyperpolarization work); this cell, along with a corresponding 10" cell (also filled with tap water) was compared via low-field NMR (SI Fig. \ref{figure:etacell}).  

Spectra were manually phased in Prospa software, but significant deviations from linear baselines required the use of Microcal Origin Pro software for processing and analysis. First, the relevant spectral regions were cut from the full 100 kHz range of data, allowing the main peak to be viewed and analyzed more specifically. Each spectrum (from each cell, with and without apodization) was reprocessed with polynomial baseline fitting / subtraction and subsequent integration (with varying integration limits); integrals were also calculated from Lorentzian fits. Taken together, the resulting values from several iterations were then used to calculate an average integral and uncertainty (standard deviation) for the 2" and 10" cells, allowing $\eta_{cell}$ to be calculated from the ratio of these two values as reported in Table \ref{tab:master} (see figure caption for SI Fig. \ref{figure:etacell}).

\subsection{Determination of $\boldsymbol{\eta}_{\boldsymbol{T}}$}

\begin{figure}[h]
    \centering
    \includegraphics[width=0.8\textwidth]{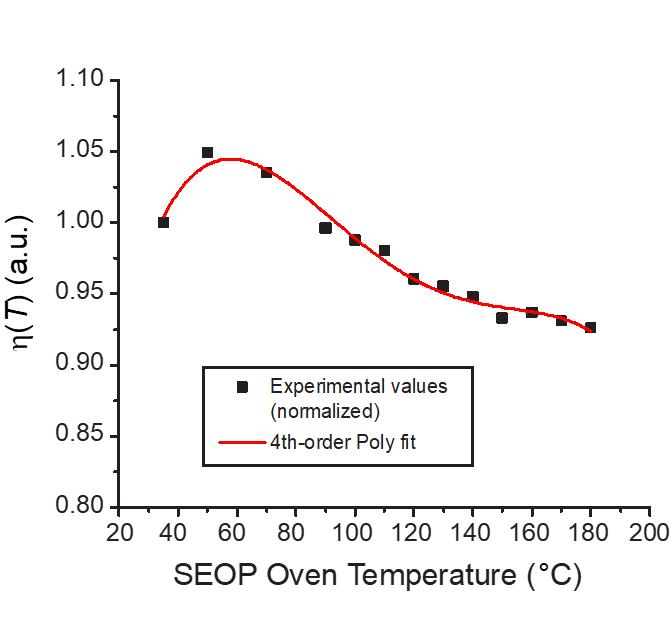}
    \caption{Graph showing $\eta_T$ values for each temperature, which were calculated using normalized integrals of signals detected by the surface coil, created by a waveform generator and an untuned broadcast coil mounted above the SEOP oven. A 4$^{\mathrm{th}}$ order polynomial fit was applied to these values (red curve; adjusted R-square value is 0.97158). Values for the correction term $\eta_T$ for Equation \ref{eq:master} are taken from the curve; the relative error value for this data was calculated by taking the root mean square of the differences between the data points and the fitted curve. The final error values were calculated by multiplying the fitted curve data points by this relative error.}
    \label{fig:eta_T}
\end{figure}

Another potential source of systematic error that may need to be considered for a given experiment is the  temperature dependence of the detection circuit, given the difference in temperature of the thermally polarized (water) and hyperpolarized (xenon) samples.  While no significant difference was previously noted over the narrower range of SEOP oven temperatures encountered in HP $^{129}$\/Xe experiments (reported elsewhere, e.g. Ref. \cite{nikolaou_maps_2014}), the significance of any effect was re-checked here given the much larger range of temperatures employed for $^{131}$\/Xe SEOP. 

The normalized integrals of the simulated NMR data created using a signal generator are plotted in SI Fig. \ref{fig:eta_T} (35 $^\circ$C is used as the reference point; this is the lowest temperature point, as the data were acquired while the laser was on and the oven air supply was flowing). The data shown are comprised of two different data sets (over two different overlapping temperature ranges; points taken at identical temperatures were averaged).  The data indicate a weak but measurable dependence of the LCR circuit on temperature over the relevant range. A 4$^{\mathrm{th}}$ order polynomial fit to the data was performed in Origin Pro; this curve was used to provide values and uncertainties for $\eta_T$ for Eq. \ref{eq:master}.  As expected, values were near unity over the relevant temperature range.

\subsection{Determination of $\boldsymbol{\eta}_{\boldsymbol{f}}$}

\begin{figure}[h]
\centering
    \includegraphics[width=0.8\textwidth]{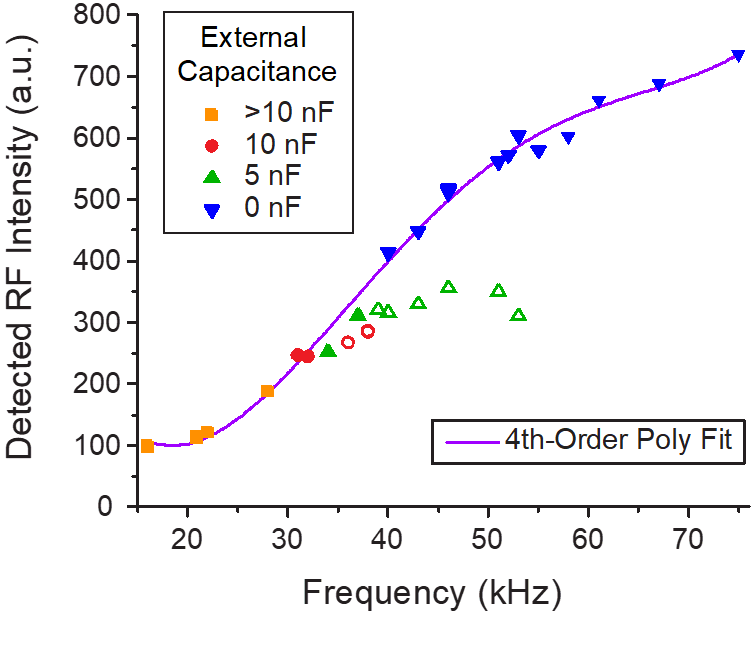}
    \caption{Signal intensity values obtained as a function of frequency for the Aurora low-field NMR spectrometer detection circuit; simulated NMR signals were provided by an RF waveform generator with constant amplitude output, broadcasted by an untuned coil mounted above the SEOP oven. All data obtained at room temperature. For each frequency, the spectrometer was manually tuned by varying the circuit capacitance; fine tuning was achieved by varying the internal capacitance (using the programmable array built into the spectrometer), whereas coarse tuning was achieved by manually varying additional circuit capacitance external to the spectrometer (see legend).  Data points obtained for configurations that were actually optimized and used for NMR detection are shown with filled icons; additional points are shown with open icons.  A 4$^{\mathrm{th}}$ order polynomial was fit to the former set (data points with filled icons) to provide $\eta_f$, calculated as the ratio between curve values obtained at the frequencies at which HP and REF spectra were acquired.}
    \label{fig:eta_f}
\end{figure}

As shown above, $^1$\/H NMR reference spectra could be reliably obtained at or above 46 kHz, allowing absolute $P_{131Xe}$\/ values for $^{131}$\/Xe NMR spectra obtained at 46 kHz to be calculated directly (i.e. for Dataset(2)).  However, for the earlier dataset obtained near 21 kHz (Dataset(1), wherein $^{131}$\/Xe values were known to be almost certainly higher because of the better conditions in the cell with respect to the Rb, and correspondingly higher spin-up rates), such direct referencing was therefore not possible with our setup.  In principle, $P_{131Xe}$\/ values could instead be calculated using HP $^{129}$\/Xe NMR spectra obtained at both frequencies (with {\em their} absolute polarization values determined by referencing to water signals obtained at the same higher frequency (46 kHz) (e.g., Ref. \cite{kailidiss}); however, this effort proved challenging here because of (1) the low enrichment of $^{129}$\/Xe spins in the gas mixture (4.79$\%$\/); (2) the very different conditions for optimally polarizing $^{129}$\/Xe and $^{131}$\/Xe spins; and (3) the difficulty of rapidly and reproducibly re-shimming the magnet at the two different frequencies.  

Following completion of all $^{131}$\/Xe experiments, determination of absolute polarization values for Dataset(1) was enabled instead by carefully quantifying the frequency dependence of the sensitivity profile for the Aurora NMR spectrometer and its detection circuit.  
The same untuned coil and waveform-generator setup used above for determining $\eta_{T}$ was used here.  The frequency from the waveform generator was varied systematically between 16 kHz and 70 kHz (SI Fig. \ref{fig:eta_f}); at each point, the spectrometer's detection circuit was manually retuned by varying both internal and external capacitance (see SI Fig. \ref{fig:eta_f} caption for additional details).  A 4$^{th}$ order polynomial was then fit to the data to obtain a generalized frequency-sensitive relationship, allowing the correction factor $\eta_f$ to be calculated that allows conversion between any two frequencies by taking the intensity value from the curve at one frequency (REF) and then dividing it by the intensity at another value (HP). For example, the value for $\eta_f$ for converting between 46 kHz and 21 kHz is 4.68 $\pm$ 0.07.   

\newpage

\section{Supplementary $^{131}$\/Xe Data, Second Data Set}
\subsection{Examples of $^{131}$\/Xe NMR and Near-IR Spectral Data}

\begin{figure}[H]
    \centering
    \includegraphics[width=0.9\textwidth]{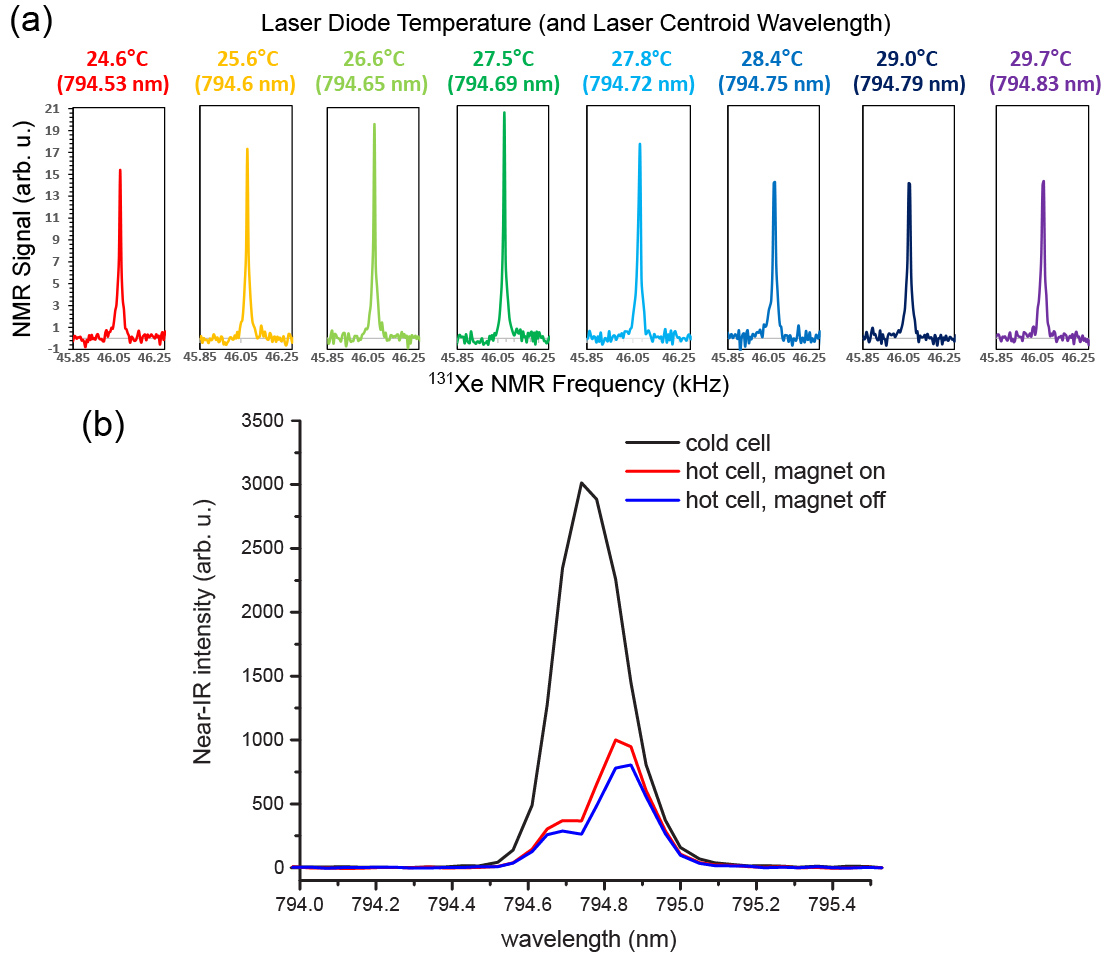}
    \caption{
    (a) Examples of low-field $^{131}$\/Xe NMR spectra taken from the second dataset (specifically, spectra used to make Fig. 3b in the main document, acquired following SEOP with the pump laser tuned to different centroid wavelengths near the Rb D1 line by varying the diode temperature). (b) Examples of near-IR spectra from the pump laser after transiting through the cell, taken from a fiber-optic probe placed behind the retro-reflection mirror (through which a small amount of light passes).  Spectra were taken with the SEOP cell either ``cold" (i.e. at room temperature, black curve), ``hot" (i.e. cell at 180 $^{\circ}$\/C), with the electromagnet ``on" (red curve); or ``hot", with magnet ``off" (blue curve). Spectra shown were taken with the laser diode temperature set to 28.3 $^{\circ}$C; the dip in the spectral lines corresponds to the center of the Rb D1 absorption. The changes in laser transmission from field cycling can be used to estimate $\langle P_{Rb} \rangle$ (here, $\sim$14.6\%) \cite{whitingJMR}.  }
    \label{fig:131andIRexamples}
\end{figure}

\newpage

\subsection{Supplementary $^{131}$\/Xe Polarization Dynamics Data}

\begin{figure}[H]
    \centering
    \includegraphics[width=0.9\textwidth]{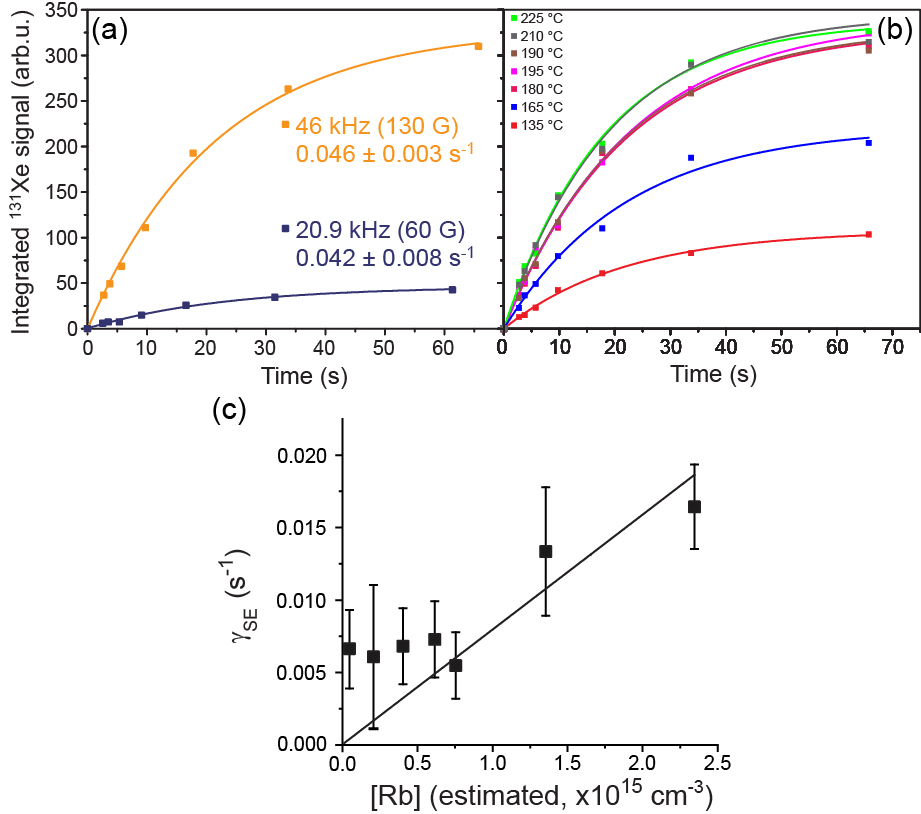}
    \caption{
(a) {\it In situ} low-field NMR showing $^{131}$\/Xe polarization build-up, with acquisitions occurring at two different magnetic fields ($\sim$\/60 and 130 G, or 20.9 and 46 kHz resonance frequency, respectively). The curves are exponential fits; the $\Gamma_{SEOP}$\/ rate constants (provided in the figure) agree within the fit error bars, showing that the build-up rates are not significantly field dependent over this range. (b) Same as (a), but as a function of SEOP cell temperature.  The corresponding plot of estimated Rb/$^{131}$\/Xe spin-exchange rates versus estimated Rb densities is shown in (c) ({\it cf.} Figs. 1 and 2 in the main document, performed using the same procedure and estimates of $^{131}$\/Xe relaxation). The data in (b,c) were taken at the end of all experiments (and end-of-life of the Rb/$^{131}$\/Xe SEOP cell. The effective spin-exchange cross-section (taken from the fit in (c)), $\gamma^{\prime}$\/=8.2($\pm$1.2)$\times10^{-18}$\/ cm$^{3}\cdot$\/s$^{-1}$, is nearly three orders of magnitude smaller than that in Fig. 2 of the main document---consistent with significant degradation of the cell (manifested by the observation of the virtually complete absence of Rb in the optical portion of the cell, with the Rb having been ``cryo-pumped" over time into the cell's side-arm that resides outside of the oven), as indicated by visual inspection of the cell after completion of all SEOP experiments.  }
    \label{fig:lianaeradynamics}
\end{figure}

\newpage

SI References

%%%%%%%%%%%%%%%%%%%%%%%%%%%%%%%%%%%%
\bibliography{ref3.bib}
\bibliographystyle{apsrev4-1}